\documentclass[twoside]{IEEEtran}
\usepackage[pdftex]{graphicx}
\graphicspath{{PDF/}}
\DeclareGraphicsExtensions{.pdf,.jpeg,.png}

\ifCLASSINFOpdf
\else
\fi
\usepackage[cmex10]{amsmath}
\usepackage[tight,footnotesize]{subfigure}
\usepackage{rotating}
\usepackage{multirow}

\begin{document}
%
% Paper title
% can use linebreaks \\ within to get better formatting as desired
\title{Reconfigurable Reflectarrays and Array Lenses for Dynamic Antenna Beam Control: A Review}

% Author names and affiliations
% use a multiple column layout for up to three different
% affiliations
\author{Sean Victor Hum, \IEEEmembership{Senior Member, IEEE}, and Julien Perruisseau-Carrier, \IEEEmembership{Senior Member, IEEE}\thanks{The authors are with the Edward S. Rogers Sr. Department of Electrical and Computer Engineering, University of Toronto, Canada, and the Adaptive Micro Nano Wave Systems Group, LEMA/Nanolab, \'{E}cole Poltechnique F\'{e}d\'{e}rale de Lausanne (EPFL), Switzerland, respectively. (email: sean.hum@utoronto.ca, julien.perruisseau-carrier@epfl.ch)}}

% Use for special paper notices
%\IEEEspecialpapernotice{(Invited Paper)}

% Make the title area
\maketitle

\begin{abstract}
Advances in reflectarrays and array lenses with electronic beam-forming capabilities are enabling a host of new possibilities for these high-performance, low-cost antenna architectures.  This paper reviews enabling technologies and topologies of reconfigurable reflectarray and array lens designs, and surveys a range of experimental implementations and achievements that have been made in this area in recent years.  The paper describes the fundamental design approaches employed in realizing reconfigurable designs, and explores advanced capabilities of these nascent architectures, such as multi-band operation, polarization manipulation, frequency agility, and amplification.  Finally, the paper concludes by discussing future challenges and possibilities for these antennas.
\end{abstract}


\section{Introduction}
\label{sec:intro}

\IEEEPARstart{T}{he need} for low-cost, reconfigurable antenna beam-forming is widespread in many existing and next-generation wireless and sensing systems.  High-gain pencil-beam or multi-beam synthesis is paramount to many systems including satellite communications, point-to-point terrestrial links, deep-space communication links, and radars.  Traditional aperture antennas such as reflectors and lenses provide a relatively low-cost and straightforward solution for achieving high antenna gain.  Their downside is that adaptive beam-steering is only possible through the use of mechanical scanning, and adaptive beam-shaping is also similarly elusive unless more sophisticated feeding systems are considered.  On the other hand, phased antenna arrays provide electronic flexibility in exciting the elements, allowing for reconfiguration and scanning of the beam pattern in real time.  The disadvantage of phased arrays, however, is their large hardware footprint, as each array element (or sub-array as the case may be) needs to be connected to a dedicated transceiver module leading to very high implementation cost.  Phased arrays also diminish in efficiency at millimeter-wave frequencies due to the use of transmission-line feeding networks which become increasingly lossy at high frequencies.

Reflectarrays and array lenses are interesting hybrids between aperture antennas and antenna arrays.  They have been studied extensively in the past 20 years due to their attractive qualities, namely their low-profile nature, ease of manufacturing, low weight, good efficiency, and overall promise as  high-gain antenna alternatives.  Recently, researchers have become interested in electronically tunable versions of reflectarrays and array lenses to realize reconfigurable beam-forming.  By making the scatterers in the aperture electronically tunable through the introduction of discrete elements such as varactor diodes, PIN diode switches, ferro-electric devices, and MEMS switches within the scatterer, the surface as a whole can be electronically shaped to adaptively synthesize a large range of antenna patterns.  At high frequencies, tunable electromagnetic materials such as ferro-electric films, liquid crystals, and even new materials such as graphene can be used to as part of the construction of the reflectarray elements to achieve the same effect.  This has enabled reflectarrays and array lenses to become powerful beam-forming platforms in recent years that combine the best features of aperture antennas and phased arrays.  They offer the simplicity and high-gain associated with their reflector / lens counterparts, while at the same time providing fast, adaptive beam-forming capabilities of phased arrays using a fraction of their hardware and associated cost.  They are also highly efficient, since there is no need for transmission line feed networks as in the case of phased arrays.

This paper reviews the development of reconfigurable reflectarray (RRA) and reconfigurable array lens (RAL) technology.  While extensive and impressive advances in reflectarray and array lens technology have been made over the past 50 years, this paper focuses on key experimental achievements that have been made in the area of reconfigurable variations of these architectures, which have been primarily made in the past decade or so.  Hence, its purpose is not to provide a review of the architectures specifically, but focus more on the mechanisms and innovation by which the architectures can be realized in reconfigurable form.  %Emphasis is given to works demonstrating reconfigurable reflectarrays and array lens experimentally.

This paper is organized as follows.  It begins by discussing the basic operation of reflectarrays and array lenses and reviewing advances in the underlying architectures in Section~\ref{sec:background}.  Then, the paper introduces the underlying technologies for enabling reconfigurability in Section~\ref{sec:enabling}.  Section~\ref{sec:basic} presents basic concepts for introducing reconfigurability to reflectarrays, focusing on single-band, single-polarization beam-scannable reflectarrays.  This discussion progresses to more advanced concepts presented in Section~\ref{sec:advanced}, which presents implementations providing dual-band operation, dual-polarization capability, frequency agility, and other unique capabilities.  Reconfigurable array lenses, and their close relation and similarity in operating principles to the reflectarray, are discussed in Section~\ref{sec:arraylens},  The paper includes with a discussion of a number of future challenges to the field in Section~\ref{sec:ongoing}, to inspire readers about research that lies ahead.  Finally, conclusions are drawn in Section~\ref{sec:conclusions}.

\section{Reflectarray and Array Lens Background and History}
\label{sec:background}

Reflectarrays and array lenses originally evolved as independent architectures for approximating the behavior of reflector and lens antennas, respectively.  Here, the basic history and operating principle of each architecture is described briefly.

\subsection{Reflectarray Principles and Development}
\label{sec:raback}

\begin{figure}[htbp]
  \centering
  \subfigure[Reflectarray]
  {\includegraphics[scale=0.65]{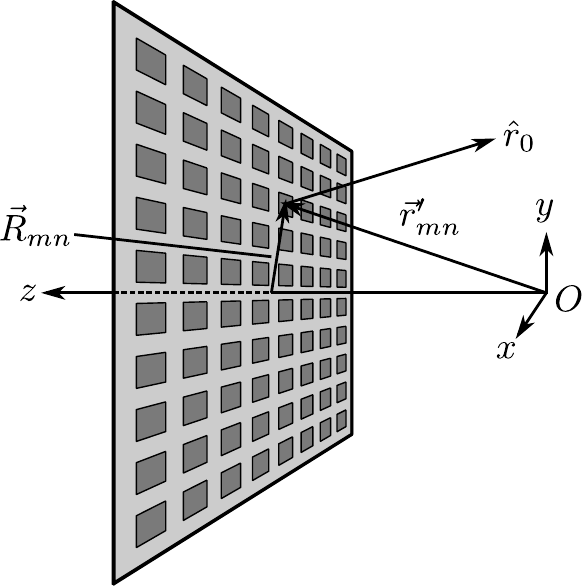}}
  %\hspace{5mm}
  \subfigure[Array lens]
  {\includegraphics[scale=0.65]{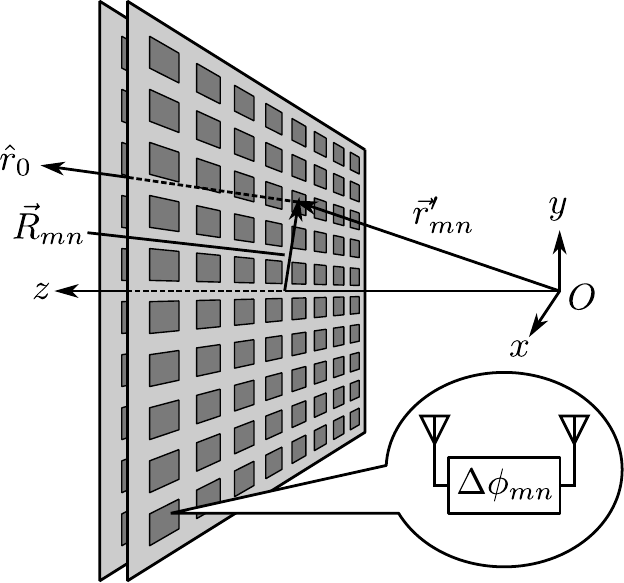}}
  \caption{Spatially-fed array architectures}
  \label{fig:architectures}
\end{figure}

The reflectarray concept was first developed by Berry in the 1960s, and utilized short-circuited waveguide sections to compensate for the phase shifts needed to collimate waves from a feed antenna into a pencil beam~\cite{berry1963}.  Interest in reflectarrays did not really begin in earnest, however, until planar antennas (namely, microstrip patch antennas) were popularized in the 1990s, which is when most advances in reflectarrays began to be made~\cite{huang1991}.  Hence, the discussions in this paper are most concerned with a planar reflectarray, which is illustrated in Figure~\ref{fig:architectures}(a).

A basic reflectarray collimates waves from a nearby feeding antenna into a pencil beam by applying a phase correction to the scattered field at each element on the reflectarray surface.  For the case of a reflectarray with a feed whose phase center is located at the origin $O$ as shown, the phase of the scattered field from the entire reflectarray must be constant in a plane normal to the direction $\hat{r}_0$ of the desired beam so that,
\begin{equation}
  k_0(r_{mn}' - \vec{R}_{mn} \cdot \hat{r}_0) - \Delta\phi_{mn} = 
  2\pi N,
  \label{eqn:beamforming}
\end{equation}
where $k_0$ is the wavenumber in free space, $\vec{r}_{mn}'$ is the position vector of the $mn$th element, $\vec{R}_{mn}$ is a position vector of the $mn$th element relative to $(0, 0, f)$, $f$ is the focal length, $\vec{r}_0$ is the desired direction of the pencil beam  and $N = 0, 1, 2, \ldots$  A phase shift $\Delta\phi_{mn}$ is introduced between the incident and scattered field by the $mn$th reflectarray element.  

However, it is important to point out that reflectarrays can do more than synthesize pencil beams.  They are popular options for contoured-beam synthesis as well as multi-feed systems, for which more advanced design methods must be pursued.  Additionally, fast vectorial analysis techniques allow for the prediction of cross-polarization, the effect of varying the angle of incidence, and so on~\cite{huang2007book}.

Most of the design effort in reflectarrays has been in realizing suitable fixed elements that synthesize the desired phase shift as some part of the element's geometry is varied.  These elements must provide a large range of phases to accommodate the geometry of the reflectarray, and the phases must be as linear with frequency as possible, if good bandwidths are to be achieved.  Additionally, the magnitude of the scattered wave is ideally the same as that of the incident wave.  Steady research progress on reflectarrays has allowed design and analysis techniques for the structures to mature significantly in recent years~\cite{huang2007book}.  Fast, fully vectorial analysis techniques enable one to predict attributes such as cross-polarization performance, the effect of varying angles of incidence on the elements, and so on.

Most reflectarray designs in the literature present a variety of designs for fixed reflectarrays whereby the $\Delta\phi_{mn}$ terms are static.  Linearly-polarized designs can be realized by varying the shape and size of patch elements~\cite{pozar1997}, slots~\cite{chaharmir2003}, loops~\cite{bialkowski2008}, and other element shapes~\cite{bozzi2003}.  Elements can also be coupled to transmission line stubs of varying lengths to vary the scattered phase~\cite{chang1995}.  In circularly-polarized (CP) designs, both of these approaches can be utilized by acting on the scattered phase of each polarization independently.  There also exists a third option for CP designs, whereby the element can be physically rotated to directly manipulate the phase shift~\cite{huang1998}\cite{martynyuk2004}.  %Hence, the final design consists of mask(s) that are used to pattern metallic cladding(s) on dielectric substrate(s) into the desired array of scatterers.  

Many reflectarray elements capitalize on resonances in the scatterer to achieve the large phase shift between the incident and scattered waves.  Hence, the effect tends to be narrowband and much of the recent research on fixed-beam reflectarrays has been devoted to realizing broadband, or multi-band, designs.  While a complete list would be too long to present here, approaches to achieve wideband element designs tend to focus on either coupling multiple resonances together~\cite{encinar2001}--\nocite{encinar2003}\cite{chaharmir2010} or coupling antenna elements to true time delay (TTD) lines~\cite{carrasco2008}.  Multi-band designs are also similarly achieved by stacking multiple resonators together~\cite{hsu2007}, or overlaying resonators on the same metal layer~\cite{yu2010}.  Most recently, the use of sub-wavelength elements has been identified as an effective means for improving reflectarray bandwidth~\cite{pozar2007}.  This essentially makes the reflectarray look more like an artificial impedance surface~\cite{sievenpiper1999} whose localized reflection coefficient can be controlled over a larger bandwidth~\cite{nayeri2010awpl}\cite{edalati2012}.  As we will see in Section~\ref{sec:basic}, the impedance surface concept is not dissimilar from modern wideband implementations of reflectarrays.

\subsection{Array Lens Principles and Development}
\label{sec:alback}

Array lenses, also known as constrained lenses and transmitarrays, were first realized by controlling the delay of an electromagnetic wave as it passed through a discrete structure~\cite{kock1949}.  They attracted significantly more interest once planar antenna technologies were available, and waves could be coupled to delay lines connecting the input and output array elements composing the array lens~\cite{mcgrath1986}.  Microstrip elements were very popular for exploring early array lenses~\cite{pozar1996}, though parallel efforts, while not strictly array lenses, were extensively investigated in the context of spatial power combiners~\cite{popovic1998}.  A schematic of an array lens is shown in Figure~\ref{fig:architectures}(b).  %Although many implementations of array lenses are possible, the schematic shows an input array coupled to an output array by some phase-shifting mechanism.

%\begin{figure}[htbp]
%  \centering
%  \includegraphics[scale=1]{arraylens}
%  \caption{Array lens schematic}
%  \label{fig:arraylens}
%\end{figure}

Similar to reflectarrays, the goal of an array lens is to typically to collimate waves from a feed into a pencil beam on the output side of the lens.  Hence, the beam-forming equation (\ref{eqn:beamforming}) is the same, except that the desired pencil beam appears on the opposite side of the surface as the reflectarray shown in Figure~\ref{fig:architectures}(a).  A key difference in the design of array lens elements is that in addition to exhibiting a large phase range and low insertion loss, the element should produce low (ideally zero) reflection from the input side of the element.  Unlike reflectarrays, where the pencil beam can be potentially directed in the specular direction to minimize reflection losses, power is permanently lost to specular reflections in array lenses.

Originally, array lenses were conceived as the inter-coupling of antenna elements on the input side of the lens to corresponding elements on the output side of the lens, as shown in the inset of Figure~\ref{fig:architectures}(b).  The simplest phasing mechanism of array lenses is a length of transmission line chosen for the required phase shift~\cite{pozar1996}\cite{barba2006}.  However, in principle any two-port network can be used to provide the phase shift provided it can be encapsulated within the array lens. %, including amplifiers.  
%In fact,  insert power amplifiers between the input and output, though generally the length of the transmission lines is chosen to re-focus the output of the combiner at a point in space where the amplified waves can be collected, rather than chosen to phase the output elements to radiate into a pencil beam in the far-field.

The phasing network does not necessarily need to be a guided-wave transmission line circuit.  The input and output antenna elements can be coupled via other microwave structures, such as slots, which can be patterned to provide a specific frequency response~\cite{abbaspour-tamijani2004}, including potentially the phase shift.  Additionally, phase-shifting of circularly-polarized radiation from the feed can be accomplished using element rotation~\cite{phillion2011}.  Furthermore, similar to reflectarrays, array lenses can be composed of resonant scatterers that couple together to impose the required phase shift on the incident wave~\cite{milne1980}\cite{ryan2010}.  Essentially, the array lens becomes a nonuniform frequency selective surface (FSS) when realized in this way, except that the local insertion phases of the elements become the primary design objective, rather than the overall magnitude response (filtering effect) of a fully periodic FSS~\cite{al-joumayly2011}.  

Examining (\ref{eqn:beamforming}), it can be readily seen that the phase shift $\Delta\phi_{mn}$ could be adaptively controlled in order to provide dynamic beam-forming or beam-synthesis capabilities from reflectarrays and array lenses alike.  This potential capability in reflectarrays was identified early on in their development~\cite{cooley1997} as a significant advantage.  In the next section, tunable technologies that enable this reconfigurable phase shift are presented, and the subsequent sections will provide specific details on how a wide variety of adaptive beam-forming platforms can be realized from these technologies.

\section{Enabling Reconfiguration Technologies}
\label{sec:enabling}

There are various enabling technologies for the dynamic control of electromagnetic waves in RRAs and RALs, which differ significantly in terms of maturity, availability, performance, or other characteristics such as integration and biasing complexity, or the suitability to a given frequency range. Therefore it is crucial to select the best technology for a given implementation and set of requirements. Though a detailed review on reconfiguration technologies is beyond the scope of this paper, it is important here to overview the main solutions available to the antenna designer and highlight their key properties regarding RRA and RAL implementations. 

There has been significant progress in the development and application of reconfiguration technology platforms for antennas and other microwave devices in recent years, mainly driven by the increased demand for adaptability or multi-functionality in radar and communication systems.  As a result emerging technologies have been consolidated (e.g.~MEMS) and exotic solutions recently introduced, such as photo-conductive~\cite{chaharmir2006tap}, macro-mechanical~\cite{Leg2009a}, fluidic~\cite{Lon2011a}, and graphene-based~\cite{Car2013a} reconfiguration techniques.  Table~\ref{tab:technologies} provides an overview of the main properties and suitability of the technologies.  Other criteria such as power handling and required control voltage also have to be considered in practice.  It is important to emphasize that different entries in the table are not always independent and should be regarded as general qualitative assessment; in practice the definition of a specific application and requirements for a specific RRA or RAL design would allow a more accurate selection of the optimal technology.

\begin{table}[htbp]
  \centering
  \caption{Selected technologies for the implementation of RRA and RALs and qualitative assessment of a few related properties (`+' , `0', and `-' symbols refer to good, neutral, and poor, respectively).}
  \begin{minipage}[c]{\columnwidth}
    \renewcommand{\thempfootnote}{\arabic{mpfootnote}}
  \begin{tabular}{|l|l|l|l|l|l|l|l|l|}
    \hline
    Type & Technology & \begin{sideways}Maturity - reliability\end{sideways} &
    \begin{sideways}Integration (incl.\ biasing)\;\end{sideways} &
    \begin{sideways}D/A control\end{sideways} &
    \begin{sideways}Complexity (cost)\end{sideways} &
    \begin{sideways}Loss (microwave / THz)\end{sideways} &
    \begin{sideways}Bias power consumption\end{sideways} &
    \begin{sideways}Linearity\end{sideways} \\
    \hline
    \multirow{3}{0.4in}{Lumped elements} & p-i-n diodes & + & - & D & + & -/- & - & 0\\
    \cline{2-9}
    & Varactor diodes & + & - & A & + & -/- & + & -\\
    \cline{2-9}
    & RF-MEMS & 0 & + & D\footnote{While analog MEMS is possible, digital MEMS devices have been proven to be more reliable / repeatable.} & + & +/0 & + & +\\
    \hline
    Hybrid & Ferro-electric & 0 & + & A & 0 & 0/- & + & 0\\
    & thin film & & & & & & & \\
    \hline
    \multirow{4}{0.4in}{Tunable materials} & Liquid crystal & 0 & 0 & A & 0 & -/+ & 0 & 0\\
    \cline{2-9}
    & Graphene & - & + & A & 0 & -/+ & + & -\\
    \cline{2-9}
    & Photo-conductive & 0 & - & A? & 0 & -/- & - & -\\
    \cline{2-9}
    & Fluidic & 0 & - & A & 0 & 0/+ & + & 0\\
    \hline
  \end{tabular}
  \label{tab:technologies}
\end{minipage}
\end{table}

%\pdfannot width 1in height 1in depth 1in
%{ /C [0.2 0.2 0.2] /Subtype /Text /Contents (text) }

The solutions in Table~\ref{tab:technologies} are classified according to whether the control is made using variable lumped element to be embedded in the array unit cell, or via the distributed control of some material property. 
Most designs so far use lumped elements, and in particular semiconductors elements such as $\textrm{p-i-n}$ and varactor diodes~\cite{hum2007}\cite{Kam2011a}. This is mainly due to the maturity and availability of off-the-shelf components, but also to the fact that this technology does not require advanced fabrication facilities or expertise. To overcome the well-known limitations of such technologies, RF-MEMS technology was employed~\cite{Leg2003a}--\nocite{hum2006}\cite{perruisseau-carrier2008}, the most prominent properties of which being very low loss up to mm-wave frequencies, virtually zero power consumption, high linearity, and possibility of monolithic integration. One limitation of MEMS technology for RRAs and RALs is that analog control generally does not provide sufficient reliability or temperature stability, and thus two-state digital elements are used, similar to the use of $\textrm{p-i-n}$ diodes in semiconductor technology. This implies increased unit cell and biasing network complexity.  Ferroelectric thin-films have also being used to implement RRAs~\cite{Rom2007a}. This technology has the advantage of providing analog control in a monolithic fabrication process and using very low power. However, losses quite higher than those achievable with MEMS. 

The DC biasing network is a particularly acute issue in RRA and RALs, since in general each cell of the array must be controlled independently, potentially resulting in thousands of control lines. Technologies offering a maximum of 1 bit control per lumped element such as $\textrm{p-i-n}$ diode and most RF-MEMS technologies will result in a larger number of biasing commands, resulting in a tradeoff between performance and complexity when selecting the elementary phase resolution. This issue is related to the well-known phase quantization effects in antenna arrays~\cite{Mai2005a}: phase errors made at each element due to the finite number of available phase states result in reduced gain and rising side lobe levels. For this reason, in large arrays it might be interesting to consider phase resolution of reflective elements as low as 1-bit~\cite{Eba2009a}--\nocite{Deb2013a}\cite{montori2010}. In any case, the biasing network has to be carefully designed not to affect the device and scattering performance. In this regard, it is important to note that advanced MEMS processes readily include highly resistive layers allowing realizing very high impedance bias line transparent to the EM waves, which is extremely convenient for the biasing network design.

Though MEMS is becoming a mature technology and can provide excellent properties up to V or W band, new technologies are still needed to address the growing interest in mm-wave and THz frequencies for communication and sensing. This issue is especially relevant for RRAs and RALs, whose space-feeding is essential for reducing loss in feeding of the array element as frequency increases. In this context recently liquid crystal (LC) technology has been considered for sub-millimeter-wave frequencies~\cite{perez-palomino2013}.  It has been proposed to address upper terahertz or even infrared frequencies using graphene~\cite{Car2013a}\cite{Car2013b}.  Interestingly, these emerging technologies allow simple biasing via a single electrode per cell since the material properties are controlled in an analog fashion.

Another important aspect when comparing lumped element and tunable material technologies for the design of RRA or RAL cells concerns modelling and design. In particular, the design of a lumped elements based cell can be carried out representing it by a multi-port scattering matrix where the effect of the lumped elements is included via circuit-based post-processing. This not only allows a single full-wave simulation of the cell for obtaining all the different states of the cell~\cite{perruisseau-carrier2010mtt}, but also allows for other interesting analyses such as the average or maximum voltage induced on each element~\cite{perruisseau-carrier2010eucap} or some computation related to the sensitivity of the cell response to faults in the lumped control devices~\cite{salti2010}. However it is worth noting here that accurate results require rigorous correction of parasitics related to the introduction of the lumped port in the full-wave simulator~\cite{perruisseau-carrier2010mtt}\cite{You2012a}. Obviously, this separate computation of cell response and control elements is virtually impossible for technologies relying on the distributed control of some material property, which thus require full-wave solutions for each material state and provide fewer possibilities for advanced optimization methods.

\section{Basic Reconfigurable Reflectarray Approaches}
\label{sec:basic}

\begin{figure}[htbp]
  \centering
  \subfigure[Tunable resonator]
  {\includegraphics{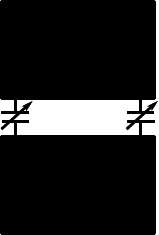}}
  \hspace{1cm}
  \subfigure[Guided-wave]
  {\includegraphics{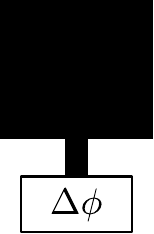}}
  \hspace{1cm}
  \subfigure[Element rotation]
  {\includegraphics{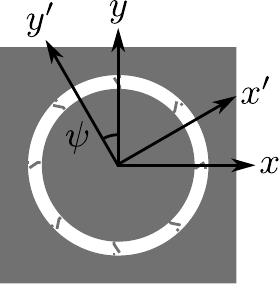}}
  \subfigure[Varactor-tuned resonator~\cite{hum2005}]
  {\raisebox{5mm}{\includegraphics[scale=0.37]{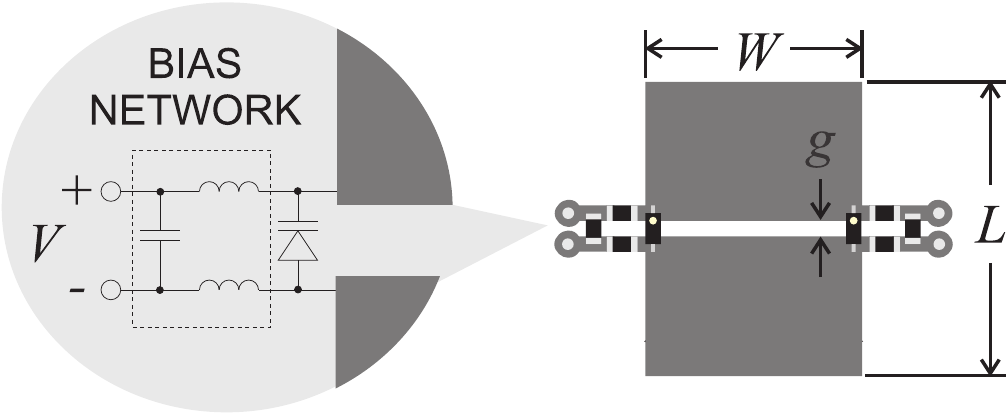}}}
  \subfigure[Tuned stub~\cite{riel2007}]
  {\raisebox{3mm}{\includegraphics[scale=0.375]{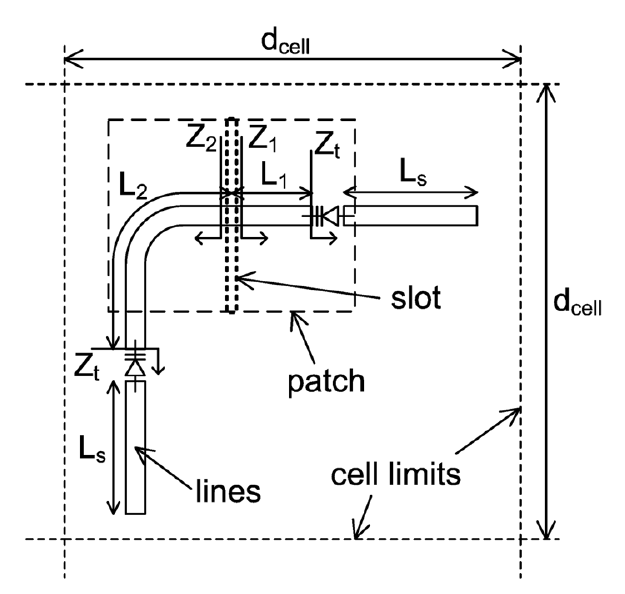}}}
  \subfigure[Spiraphase~\cite{rodriguez-zamudio2012}]
  {\includegraphics[scale=0.4]{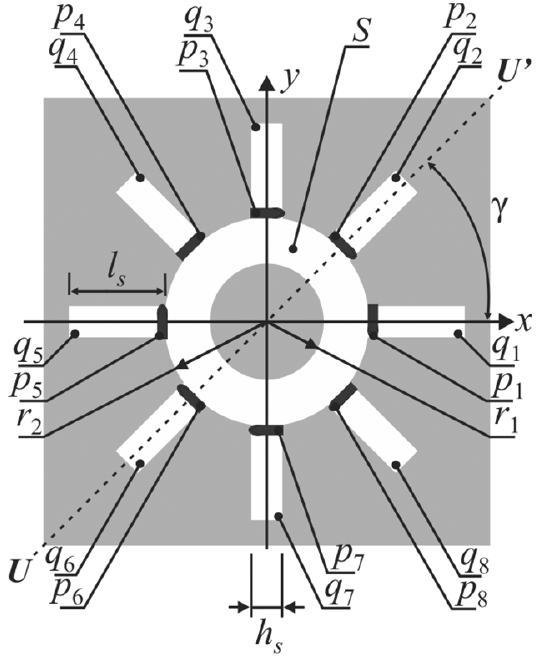}}
  \caption{Reconfigurable reflectarray approaches and corresponding examples}
  \label{fig:approaches}
\end{figure}

There are three general approaches employed in the design of basic reconfigurable reflectarrays, which are summarized in Figures~\ref{fig:approaches}(a)-(c).  Here, we define a basic reflectarray design as one operating at a single frequency on a single polarization; more advanced designs will be considered in Section~\ref{sec:advanced}.  The majority of reflectarray designs in this category manipulate the phase of the scattered field from the elements by changing characteristics of a resonator composing the elements.  One of many possible approaches is shown in Figure~\ref{fig:approaches}(a), whereby a tunable capacitor is integrated with the resonator.  Hence, if an electronically tunable phase shift is desired, a tuning mechanism can be incorporated into the resonators to make this possible.  It is also possible to evoke a phase shift from the element by transitioning received space-waves by the element to guided-waves, phase-shifting the wave using a guided-wave circuit such as a transmission line stub, and then re-radiating the resulting wave.  This approach is shown in Figure~\ref{fig:approaches}(b).  Hence, to make the phase shift dynamic electronic phase-shifting circuits can potentially be employed in the guided-wave portion of the element, resulting in an antenna / phase-shifter / antenna signal flow.  Finally, for CP waves, electronic means for element rotation can be considered to produce the necessary phase shifts, as shown in Figure~\ref{fig:approaches}(c).  Each of these three techniques are elaborated upon in more detail in the following sections.

\subsection{Tunable Resonator Approach}
\label{sec:resonator}

%Reflectarrays operating in a single, narrow frequency band often employ reflectarray elements based on resonators.  
While fixed reflectarrays modify the resonator dimensions to change their resonant frequency, and hence phase shift, reconfigurable elements achieve this using electronic tuning.
%Incident waves couple to the resonator based on the resonator's geometry, allowing the resonator to play a role in manipulating the scattered field.  The basic operating principle is that by changing the resonant frequency of the resonator, large changes in the phase of the scattered field from the resonator can be achieved.  Simple resonators, such as patches and dipoles, can be printed on a grounded dielectric slab and the resonator length changed in order to change the resonant frequency~\cite{pozar1997}.    
%
%In the case of patch elements operating in the TM$_{010}$ mode, it is the resonant length of the patch that is responsible for most of the phase change, though the width of the patch also influences the degree the degree of coupling to the resonator and hence plays some role the shape of the phase curve~\cite{karnati2013}.  
%
Electronic means for changing the resonant frequency of patches have been known for a long time, for example, through the use of frequency-agile patches employing varactor diodes~\cite{bhartia1982}, and hence the first electronically tunable reflectarray element was based on this frequency-agile patch design~\cite{boccia2002_ap}.  However, it is important to properly couple the choice of the tuning element to the size of the patch in order to achieve the large phase ranges achievable with comparable fixed elements, and this early design only achieved about $180^\circ$ of phase range.  More phase range was achieved from this varactor-loaded patch concept by contemplating different loading schemes for the patch~\cite{hum2005}\cite{vendik2008} and coupling the varactor to patches of appropriate size~\cite{boccia2010}.  It is also possible to use micro-electrical-mechanical systems (MEMS) varactors for the same purpose~\cite{hum2006}.  Figure~\ref{fig:approaches}(d) shows an example of integrating varactor diodes into the structure of a patch antenna to achieve phase agility.

Essentially, these techniques can be thought of as changing the effective electrical length of the resonator.  Hence, a wide variety of techniques have been contemplated to implement reflectarray elements based on this concept.  Switches in the form of PIN diodes and micro-electrical-mechanical systems (MEMS) have been integrated with patches to control the current path and corresponding resonator length~\cite{perruisseau-carrier2008}\cite{legay2007}\cite{rajagopalan2008}.   Such methods depend on modelling techniques that allow for the analysis of the effect of tunable lumped element devices on the large scale electrical scattering characteristics of the device~\cite{perruisseau-carrier2010mtt}\cite{aubert2006}.  In addition to using lumped element devices to effect changes in resonator lengths, more exotic techniques have also been contemplated, such as photo-induced plasmas for changing the length of slots coupled to reflectarray elements~\cite{chaharmir2006tap}.

The resonant frequency of a simple patch element also can be manipulated in a distributed fashion by varying the dielectric constant of the substrate, which is the operating principle of reflectarray elements using dielectrics with tunable properties such as liquid crystals~\cite{perez-palomino2013}\cite{moessinger2006}\cite{hu2008}.  Ferro-electric films have also been employed for in semi-distributed elements~\cite{romanofsky2000}\cite{sazegar2009}.  %The effective dielectric constant can also be manipulated by mechanically actuating the patch in the case of MEMS elements~\cite{gianvittorio2004}.  

Reflectarrays share many traits in common with artificial impedance surfaces (AISs).  Since reflectarray elements allow the phase of the scattered field to be manipulated arbitrarily, setting the phase shift to be uniform across the surface changes its electrical characteristics from that of a plain conductor.  For example, if the phase shift it set to $0^\circ$, then the reflectarray surface resembles the well-known artificial magnetic conductor~\cite{sievenpiper1999}, even though structurally the reflectarray element may be quite different from a mushroom structure.  In fact, the main differences between a reflectarray and an AIS are: i)~the dimensions of reflectarray elements are usually spaced around half a wavelength whereas in AISs the spacings tend to be smaller; ii)~the dispersion characteristics of reflectarray cells are not usually engineered to suppress surface waves; and iii)~the local phase of the reflectarray unit cells is varied in accordance with the beam to be synthesized, while AISs are fully periodic.

\begin{figure}[htbp]
  \centering
  \subfigure[AMC]
  {\includegraphics[scale=0.65]{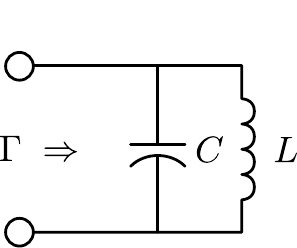}}
  \hspace{0.4cm}
  \subfigure[AIS]
  {\includegraphics[scale=0.65]{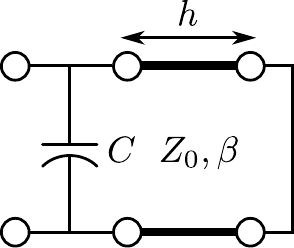}}
  \hspace{0.4cm}
  \subfigure[Reflectarray]
  {\includegraphics[scale=0.65]{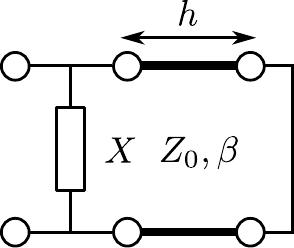}}
  \caption{Equivalent circuits of reflectarray unit cells}
  \label{fig:eqvcircuit}
\end{figure}

Equivalent circuit modelling of reflectarray unit cells also closely parallels those developed for AISs.  Each cell of a reflectarray element can be see as a scatterer placed within a periodic (Floquet) waveguide~\cite{huang2007book}.  At a specific angle of incidence, an equivalent circuit can be synthesized for the cell and the input reflection coefficient $\Gamma$ used to describe the scattering behavior of the element.  Figure~\ref{fig:eqvcircuit}(a) shows the equivalent circuit for the mushroom-style AMC which realizes a parallel LC circuit because of the intrinsic inductance of the patch/via combination and the fringing capacitance between patches.  A generalized AMC composed of, for example, floating patches can be thought of has being capacitive if the elements are sub-wavelength.  This leads to the equivalent circuit shown in Figure~\ref{fig:eqvcircuit}(b) which illustrates the equivalent capacitance of the cell placed an electrical distance $\beta h$ in front of a short-circuit, representing the ground plane on the rear of the surface.  The substrate, being illuminated by a TEM wave, acts as a transmission line, which is a typical concept from frequency selective surfaces~\cite{munk2000book}.  Finally, Figure~\ref{fig:eqvcircuit}(c) shows a possible equivalent circuit of a reflectarray element, which differs from that shown in (b) because the elements in a traditional reflectarray are typically comparable to a wavelength.  Therefore, owing to the distributed nature of the scatterer, more sophisticated circuits are needed to represent the reactance block $X$ shown in the figure~\cite{vendik2008}, or even other circuit models entirely~\cite{hum2007}\cite{karnati2013}.
  
Impedance surfaces can be easily adapted to have a tunable reflection phase.  For example, the capacitance between the patches (which appears in Figures~\ref{fig:eqvcircuit}(a) and (b)) can be made adjustable by placing a tunable capacitor such as a varactor diode across the gaps.  Tunable impedance surfaces have been demonstrated for use as plane-wave re-direction surfaces~\cite{sievenpiper2002} though the bias network can theoretically be reconfigured for such surfaces to work as tunable reflectarrays.  While the downside of this approach is that many more tunable components are needed due to the sub-wavelength size of the unit cell, the reduced unit cell size also provides for improved bandwidth characteristics~\cite{pozar2007} leading to potentially broadband reflectarray performance.  Bandwidth-related issues for tunable reflectarrays are discussed in more detail in Section~\ref{sec:ongoing}.  

\subsection{Guided-Wave Approach}
\label{sec:guided}

Rather than controlling the resonance of a scatterer as discussed so far, it is also possible to control the phase shift by a guided-wave approach, as symbolically depicted in Figure~\ref{fig:approaches}(b). In this case the incoming space-wave is first coupled by an antenna to a guided-wave. The guided-wave is then phase shifted, and is finally re-radiated, resulting in an antenna--phase-shifter--antenna topology. This technique was first applied to fixed-beam antennas, and then extended to RRAs by using dynamically controllable phase shifters as discussed in the remainder of the section. 
The guided-wave approach presents both advantages and disadvantages when compared to the tunable resonator technique. First, unit cells of the former type are generally easier to optimize. Indeed, while the modelling complexity of both approaches is quite similar (if the approach described in Section~\ref{sec:resonator} is used when addressing tunable resonator cells), the fact that the antenna and phase shifter can be optimized separately in the guided-wave approach results in simpler design procedure. For instance, in a digital design it is quite straightforward to achieve equi-spaced phase states in the guided-wave phase shifter, whereas doing so with a tunable resonator can require complex optimization which might still lead to sub-optimal phase distributions~\cite{perruisseau-carrier2010mtt}. Another advantage of this technique is that wideband behavior is more easily obtained since simple guided-wave phase shifters can be designed to provide true-time delay capability. %The bandwidth of the element is mainly limited by the antenna to phase shifter coupling mechanism.  The case of large $f/D$ arrays would require true-time delay in excess of $360^\circ$, which is problematic for both reconfigurable tunable resonator and guide wave approached because of design complexity and space for the phase shifters, respectively. 
Further comments on bandwidth of reflectarrays are provided in Section~\ref{sec:bandwidth}.

Though not strictly required, most reconfigurable guided-wave cells are implemented in multi-layer configurations~\cite{Kam2011a}, which will generally increase fabrication complexity and thermal issues. However, in some applications it might be desirable to have the tuning element shielded from the antenna aperture. 

%Finally, an important issue in the guided-wave approach is that the loss can be significantly larger than using tunable resonators. As detailed below, the loss achieved in both varactor and MEMS based guided-wave cells is much larger than the best figure obtained a tunable resonator approach. Intuitively this results from the fact that in the guided-wave approach all incoming power is flowing through the tuning circuitry (i.e. the phase shifter), while the tunable resonator approach is a more distributed control mechanism where part of the scatterer is subjected to low induced currents hence, and lower losses result.  
Several RRAs or unit cells have been developed based on the guided-wave approach, a few notable examples of which are briefly described here.  A design using antennas aperture-coupled to delay lines embedding two varactor diodes allowed achieving a continuous tuning over a $360^\circ$ range with maximum loss of 2.4 dB at 5.4 GHz~\cite{riel2007}.  Other authors proposed, as previously done in usual phased array antennas, to arrange reflectarray cells into sub-arrays to reduce the number of control elements~\cite{Car2012a}. Gathering of the elements by pairs was implemented in a full array demonstrator of 122 sub-arrays, demonstrating the possibility of cost and complexity saving without significant reduction in the performance of the antenna. Note that a similar `gathering' approach could also be used in the tunable-resonator approach, such as done previously in a Fabry-Perot antenna~\cite{Deb2010e}.

A large `guided-wave' RRA having more than 25,000 reflecting elements was fabricated for millimeter-wave imaging system operating in the 60-GHz band~\cite{Kam2011a}. To manage the complexity of this system, the unit cell for this RRA consists of microstrip patch directly connected to a 1-bit reflective transmission line embedding a $\textrm{p-i-n}$ diode.  MEMS technology has also been also considered here, and a fully-operational monolithic MEMS RRA at 26 GHz was designed and fabricated~\cite{Bay2012a}, while cells using surface mount MEMS elements were also implemented~\cite{Car2012g}. In both cases thermal losses were several dB despite the use of MEMS technology, which is below the performance that can be achieved using MEMS technology and the tunable resonator approach~\cite{perruisseau-carrier2008}.  Intuitively this results from the fact that in the guided-wave approach all incoming power is flowing through the tuning circuitry (i.e.~the phase shifter), while the tunable resonator approach is a more distributed control mechanism where part of the scatterer is subjected to low induced currents  and lower losses result.

\begin{figure}[htbp]
  \centering
  \includegraphics[width=\columnwidth]{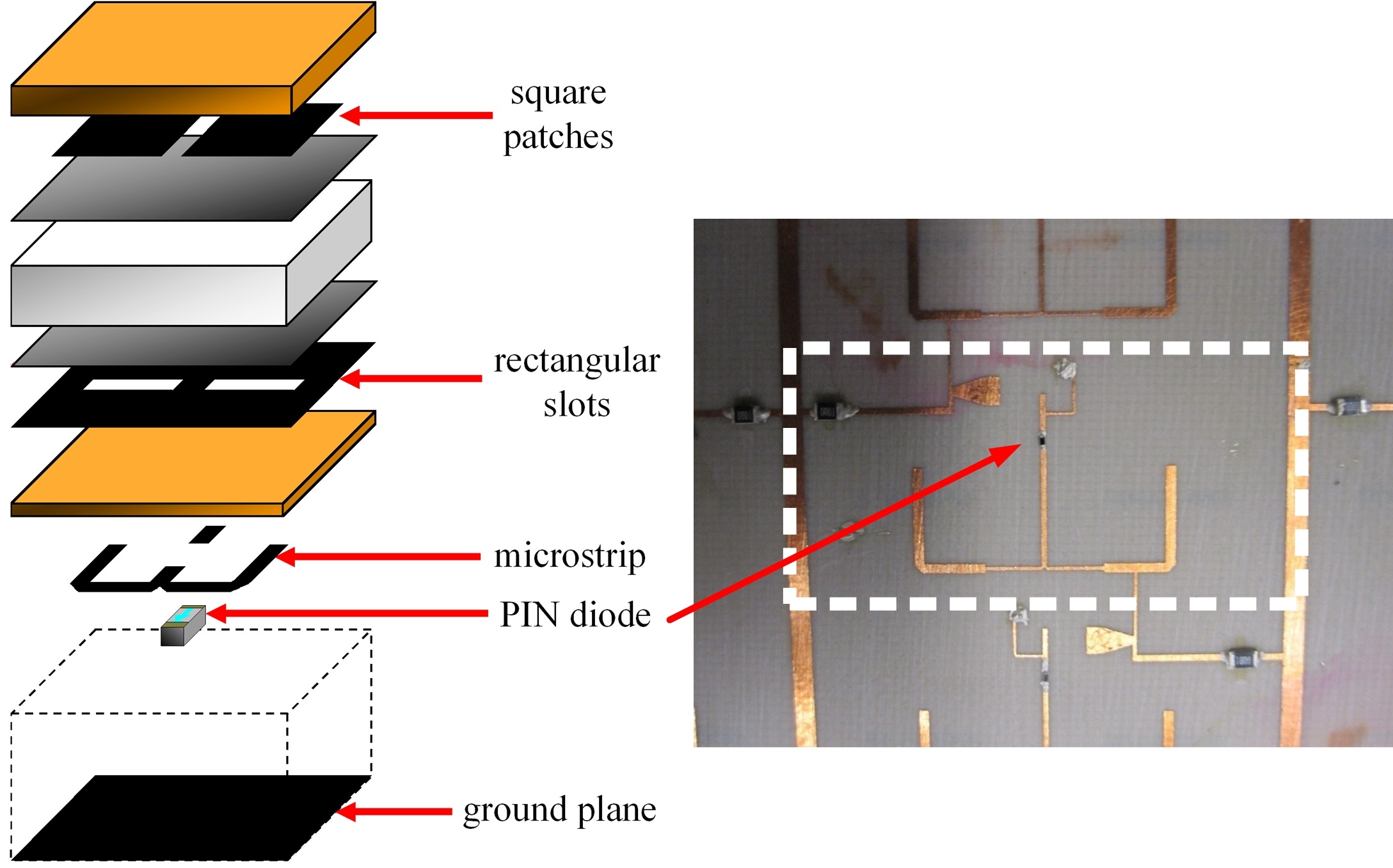}
  \caption{Illustration of the guided-wave approach to RRA phase control: patches elements are aperture coupled to a 1-bit delay line embedding a $\textrm{p-i-n}$ diode. Two antenna elements share the same phase shifter for reducing complexity~\cite{Car2012a}.}
  \label{fig:guided}
\end{figure}

\subsection{Rotation Technique for Circularly-Polarized Waves}
\label{sec:rotation}

A clever alternative to the above methods, though restricted to CP, is that of the `rotation technique'~\cite{Phe1977a}. This principle was initially applied to the reflectarray and the associated operation principle and derivations are well-known~\cite{huang1998}.  Here we summarize a slightly more general formulation for RRAs~\cite{Guc2012a} (the case of the lens array is available elsewhere~\cite{phillion2011}). 

Let us consider a general unit cell such that the unit cell is rotated an angle $\psi$ as depicted in Figure~\ref{fig:approaches}(c). Assume a right-hand-polarized feed hence an incident right-hand CP wave travelling towards the cell,
\begin{equation}
  \vec{E}^{inc} = A(\hat{a}_x + j\hat{a}_y) e^{jk_0z}.
  \label{eq:einc}
\end{equation}
It can easily be shown that the reflected field can be written in the general form
\begin{equation}
  \vec{E}^{ref} = \Gamma_{co}A(\hat{a}_x - j\hat{a}_y) e^{jk_0z} + 
  \Gamma_{xp}A(\hat{a}_x + j\hat{a}_y) e^{jk_0z}
  \label{eq:eref}
\end{equation}
with
\begin{equation}
  \Gamma_{co} = \Gamma_{co}(\psi=0)e^{j2\psi} = \left( \frac{1}{2}(s_{11}' - s_{22}') + js_{12}' \right) e^{+j2\psi},
  \label{eq:gamma_co}
\end{equation}
and
\begin{equation}
  \Gamma_{xp} = \frac{1}{2}(s_{11}' + s_{22}'),
  \label{eq:gamma_xp}
\end{equation}
where where $\Gamma_{co}$ and $\Gamma_{xp}$ are the co-polar and cross-polar CP reflection coefficients, and the primed scattering parameters correspond to the fundamental Floquet harmonics of $x'$- and $y'$-polarized waves in the primed coordinate of Figure~\ref{fig:approaches}(c). Note that even for a cell whose pattern is symmetrical around $y'$ such as in Figure~\ref{fig:approaches}(c), $s'_{12}$ is not in fact zero since a periodic arrangement of such cells along $x$ and $y$ is itself not symmetrical around $y'$. %Finally, since the array utilizes CP, here co- and cross-polarization (`co' and `xp' subscripts, respectively) refer to the circularly polarized waves having the desired and unwanted orientations, respectively.

The principle of operation and requirements for the cells are now easily deduced from (\ref{eq:einc})--(\ref{eq:gamma_xp}). First, in order to suppress the reflected cross-polarized field one must ensure that $|\Gamma_{xp}| \approx 0$, which according to (\ref{eq:gamma_xp}) requires the phase of the linear-polarized reflection coefficients along $x'$ and $y'$ axis to differ by about $180^\circ$ (assuming similar losses along both axes). In practice this is achieved by making the element resonate along $y$ at the design frequency, while being weakly excited by a $x$-oriented incident electric field. Once this condition is met, (\ref{eq:eref}) and (\ref{eq:gamma_co}) show that the phase of the desired reflected circular-polarized wave is simply twice the angular orientation $\psi$ of the element on the surface. This is the essence of the rotation technique: the reflected phase of the CP wave co-polarized wave can be simply controlled by rotating the elementary resonator along the reflector.

\begin{figure}[htbp]
  \centering
  \includegraphics[width=\columnwidth]{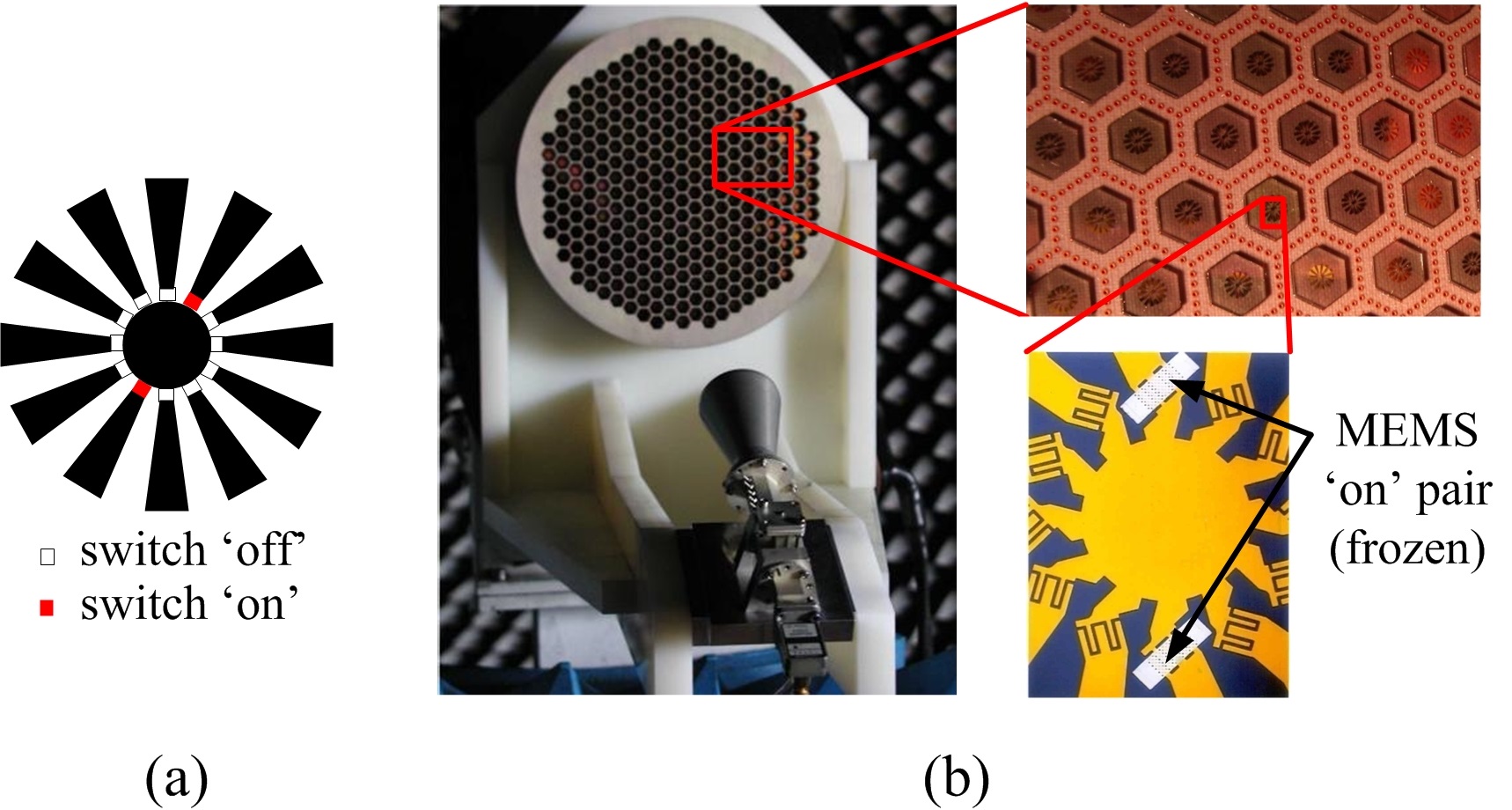}
  \caption{Reflectarray using the element rotation technique for beam-scanning for CP~\cite{Leg2003a}. (a) Example of elementary cell, (b) Array implementation using frozen MEMS states.}
  \label{fig:dioderotate}
\end{figure}

As in the case of the previous methods, the rotation technique has been implemented in fixed configurations~\cite{huang1998}\cite{phillion2011}, but has also been proposed for dynamic phase control.  In this latter case the independent rotation of each element must obviously be implemented by electrical means. 
This was proposed as early as the 1970's by integrating diodes in a rotation-invariant geometry~\cite{Phe1977a}, so that selectively actuating some of the lumped elements implements the `electromagnetic rotation' of the element.
%This has generally been achieved by integrating diode or MEMS lumped component in some rotation-invariant geometries, so that selectively actuating some of the lumped elements implements the `electromagnetic rotation' of the element~\cite{Phe1977a}\cite{Leg2003a}\cite{martynyuk2004}\cite{Guc2012a}\cite{Phi2008a}. 
An example is illustrated in Figure~\ref{fig:dioderotate}(a), where the rotation of a dipole is implemented~\cite{Leg2003a} by switching the desired pairs of branches (slots~\cite{martynyuk2004} and metal split rings~\cite{Guc2012a} have also been used).  Figure~\ref{fig:dioderotate}(b) shows a full array implementation of the concept. However, to the best of our knowledge, no operational full reflectarray with actual dynamic beam-scanning has been implemented so far, with the above examples demonstrating so-called `frozen' MEMS array implementation for complexity reasons. The use of a micro-motor for implementing the rotation has also been proposed and implemented in a unit cell~\cite{Phi2008a}, but not a full array configuration.

\section{Advanced Concepts in Reconfigurable Reflectarrays}
\label{sec:advanced}

The research on reconfigurable RRAs logically first focused on the control of a single linearly-polarized (LP) beam. These activities confirmed that the reflectarray approach is an advantageous solution in electronically-controlled antenna arrays, and motivates considering more advanced capabilities in terms of operating frequency and polarization. Specifically, the idea here is to maintain dynamic local phase control for beam-scanning/shaping, while simultaneously achieving one or several additional capabilities in terms of dual-polarization, polarization flexibility, multi-frequency, or frequency-tunable operation. Such advanced operation modes would even further the interest in reconfigurable reflectarrays, providing for instance a shared aperture for widely-spaced transmit / receive frequencies, and dual-polarization as needed in many radar and satcom applications.  Additionally, flexible frequency or polarization can be provided for cognitive radio applications~\cite{Per2010a}.

In this context it is fundamental to remark that the RRA (and to a certain extend the RAL as well) concept is inherently favorable to multi-reconfiguration when compared to standard phased arrays. This is because the implementation of more advanced control of the aperture surface comes with reduced added complexity when compared to that needed in phased array. For instance, polarization or frequency flexibility in a phased array would generally also require the implementation of reconfigurable matching networks, adding significant complexity, loss, and power consumption. Such an issue does not exist in RRAs since there is no need to match the elementary cell, which by definition reflects all non-dissipated incoming energy.  As a result various advanced RRA capabilities have been proposed recently. So far these studies essentially focused on demonstrating the capability at unit cell level, and are briefly commented on the remainder of this section.

\subsection{Dual-polarization Cells}

\begin{figure}[htbp]
  \centering
  \subfigure[Varactor-controlled unit cell for the independently beam-scanning of two single-LP]
  {\raisebox{8mm}{\includegraphics[scale=0.3]{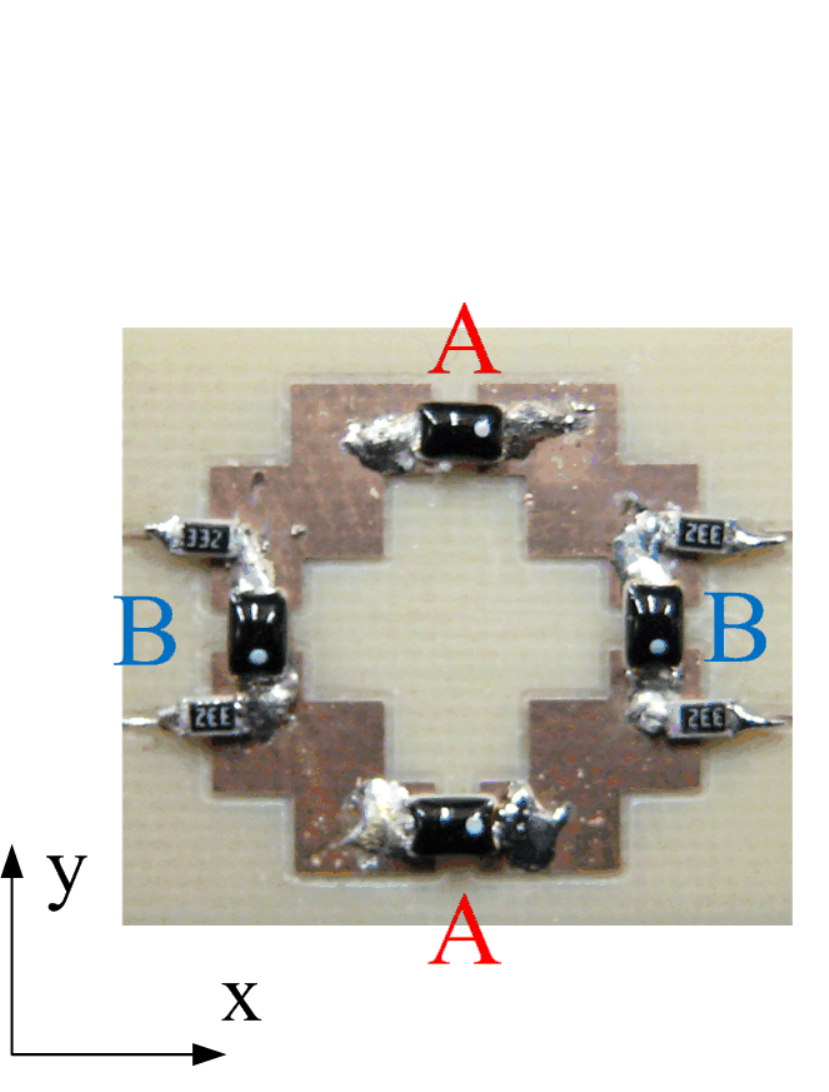}}}
  \hspace{3mm}
  \subfigure[Cell reflection phase along $x$ axis when varying the voltages of both pairs of diodes `A' and `B', demonstrating the independent polarization control]
  {\includegraphics[scale=0.3]{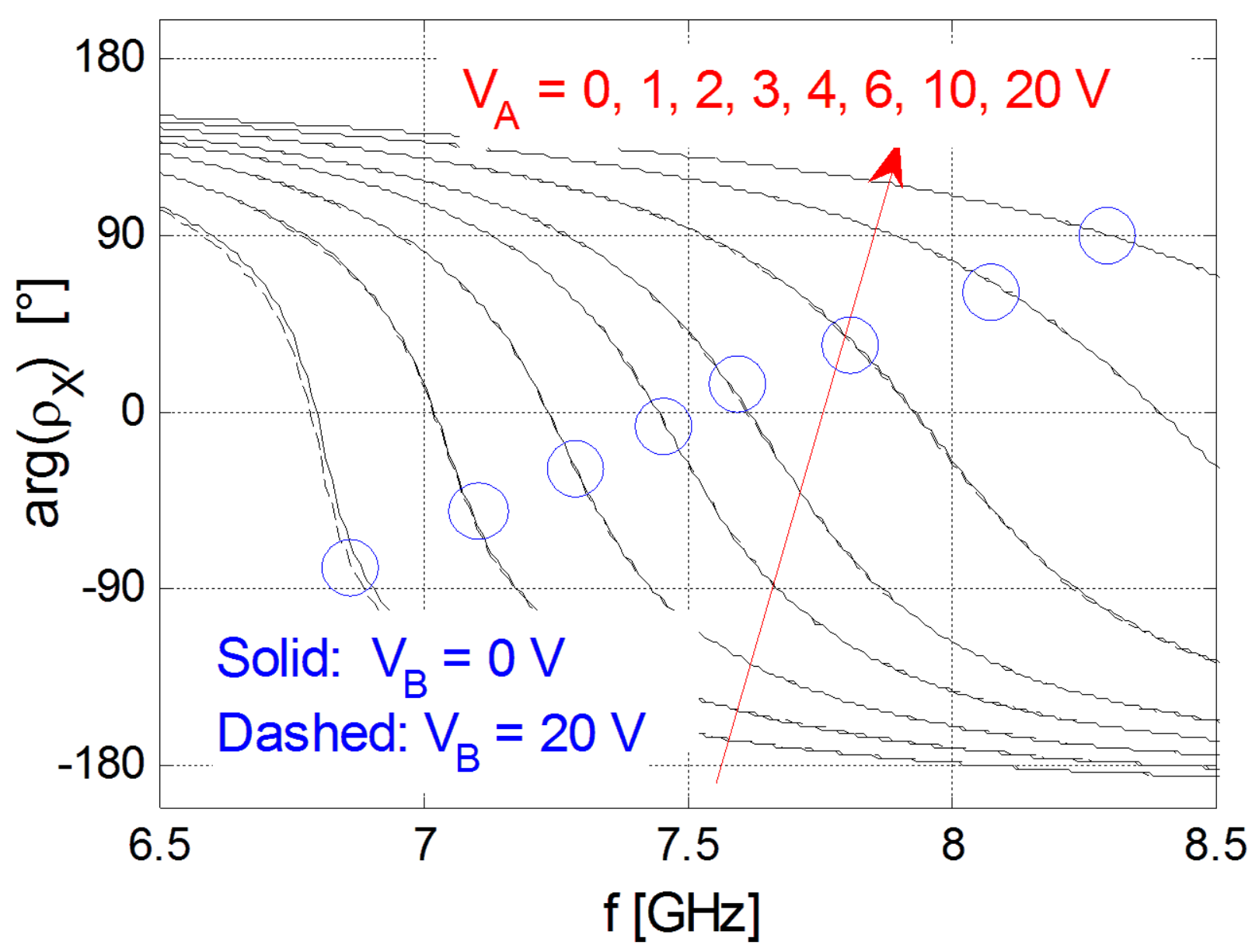}}
  \caption{Polarization reconfiguration}
  %\caption{(c) symbolic representation of the alternative operation mode of the cell for polarization-flexible single-polarization beam-scanning.}
  \label{fig:polreconf}
\end{figure}

Reflectarray cells utilizing two polarizations with independent control of the phase of each LP component, which would allow independently scanning two LP beams have been experimentally demonstrated~\cite{Per2009a},\cite{Per2010b}. The principle of such cells is illustrated in Figure~\ref{fig:polreconf}(a) and (b), where a microstrip ring resonator is loaded by two varactor diodes pairs `A' and `B'~\cite{Per2009a}.  In the case of the $y$-polarized incident field component, the varactors `B' have no effect on the reflection phase because they are located in zeros of the current distribution, by symmetry, whereas the elements `A' allow the control of the reflection phase for this polarization. In the case of the $x$ component, the control elements `A' are now in zeros of the current distribution and the reflection phase is controlled by `B'.  

The element of Figure~\ref{fig:polreconf}(a) is of the `tunable resonator approach' type described in Section~\ref{sec:resonator}. However as in the case of the single-LP cell, the dual-LP element can also be implemented using the `guided-wave approach' of Section~\ref{sec:guided}~\cite{Car2010a}~\cite{Car2013c}. In that case perfect symmetry is difficult to achieve but cross-polarization can still be made very low. In fact, the element in~\cite{Car2013c} is more robust than the initial demonstration of~\cite{Per2009a} in terms of response under oblique incidence.

\begin{figure}[htbp]
  \centering
  \includegraphics[width=\columnwidth]{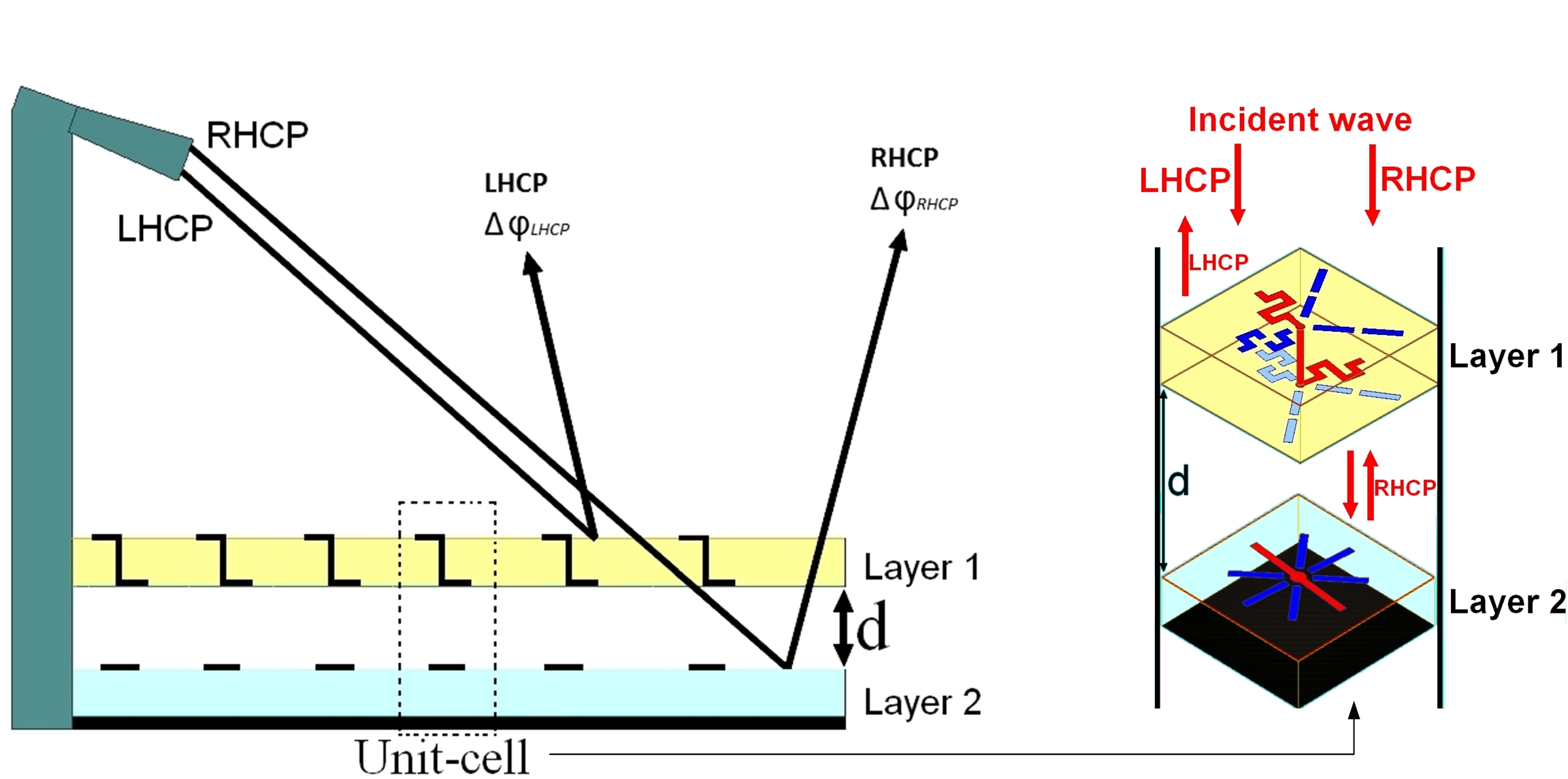}
  \caption{Principle of the dual-CP reflectarray concept of~\cite{Men2013a}}
  \label{fig:mener}
\end{figure}

More recently the implementation of a reflectarray allowing the independent control of two CP beams of opposite polarization but the same frequency~\cite{Men2013a}. Since such a capability cannot be achieved via a single-layer reflectarray, here a multi-layer structure must be adopted, as shown in Figure~\ref{fig:mener}. The top layer must be transparent to one polarization, while reflecting the other with the desired phase. The bottom layer can then be simply implemented as any single-CP reflectarray. This interesting concept has not been demonstrated experimentally yet in a true reconfigurable mode at the time of publication, but its implementation will come with similar possibilities and issues as other reflectarrays, with the additional constraint of having as many as three layers all requiring embedded control elements. 

Several applications do require dual-polarization but only with a beam common to both LP components. For instance this is the case of a line-of-sight communication between a reflectarray and a moving terminal, where the two LP components can be used as two different communication channels (so-called frequency-reuse). %In this case simpler topologies can be employed to reduce the number of control elements, as discussed in~\cite{montori2010}\cite{Per2012a}. 
Such elements logically also provide an alternative to the elementary rotation principle for single-CP reflectarrays discussed in Section~\ref{sec:rotation}.

%\textbf{What about~\cite{pereira2012}?}

\subsection{Polarization-flexible cells}

The possibility to dynamically control the polarization of the beam synthesized by a reflectarray is another very interesting prospect for cognitive radio applications, among others. In fact, cells allowing the independent control of two linear polarizations such as presented in the previous section allows achieving such a capability as well. 

%This is intuitively illustrated in Figure~\ref{fig:polreconf}(c). Consider a normally-incident single-CP wave illuminating the reflector, or similarly a single-LP wave oriented such as depicted in the figure, i.e. with $\vec{E} = E_0(\hat{e}_x + \hat{e}_y)e^{jk_0z}$.  Obviously the two degrees of freedom available allow to control the reflection phase of the wave but also its polarization, allowing LP-CP conversion or tilting of the LP orientation~\cite{Per2010b}. 
Consider the unit cell of Figure~\ref{fig:polreconf}(a) and a single LP incident field oriented such that $\vec{E}_i = E_0(\hat{e}_x + \hat{e}_y)e^{jk_0z}$. The reflected field is $\vec{E}_r = (\rho_x E_0\hat{e}_x + \rho_y E_0\hat{e}_y )e^{-jk_0z}$, where $\rho_x$ and $\rho_y$ are the reflection coefficients of the cell along the $x$ and $y$ axes, respectively. Since, as explained in Section V.A, the cell allows to \emph{independently} control $\Gamma_x$ and $\Gamma_y$,  it is possible to independently control both the polarization and the phase of $\vec{E}_r$.  This principle can be used when there is at least a 2-bit resolution for each component, since this corresponds to a $90^\circ$ phase shift step needed for conversion from LP to CP. A final important note is that high variation in the losses of the cell for different phases will strongly impact on the quality of the polarization control, hence effort must be focused on achieving similar loss in the different cell states~\cite{Car2013c}.

\subsection{Dual-band Cells}

\begin{figure}[htbp]
  \centering
  \includegraphics[width=\columnwidth]{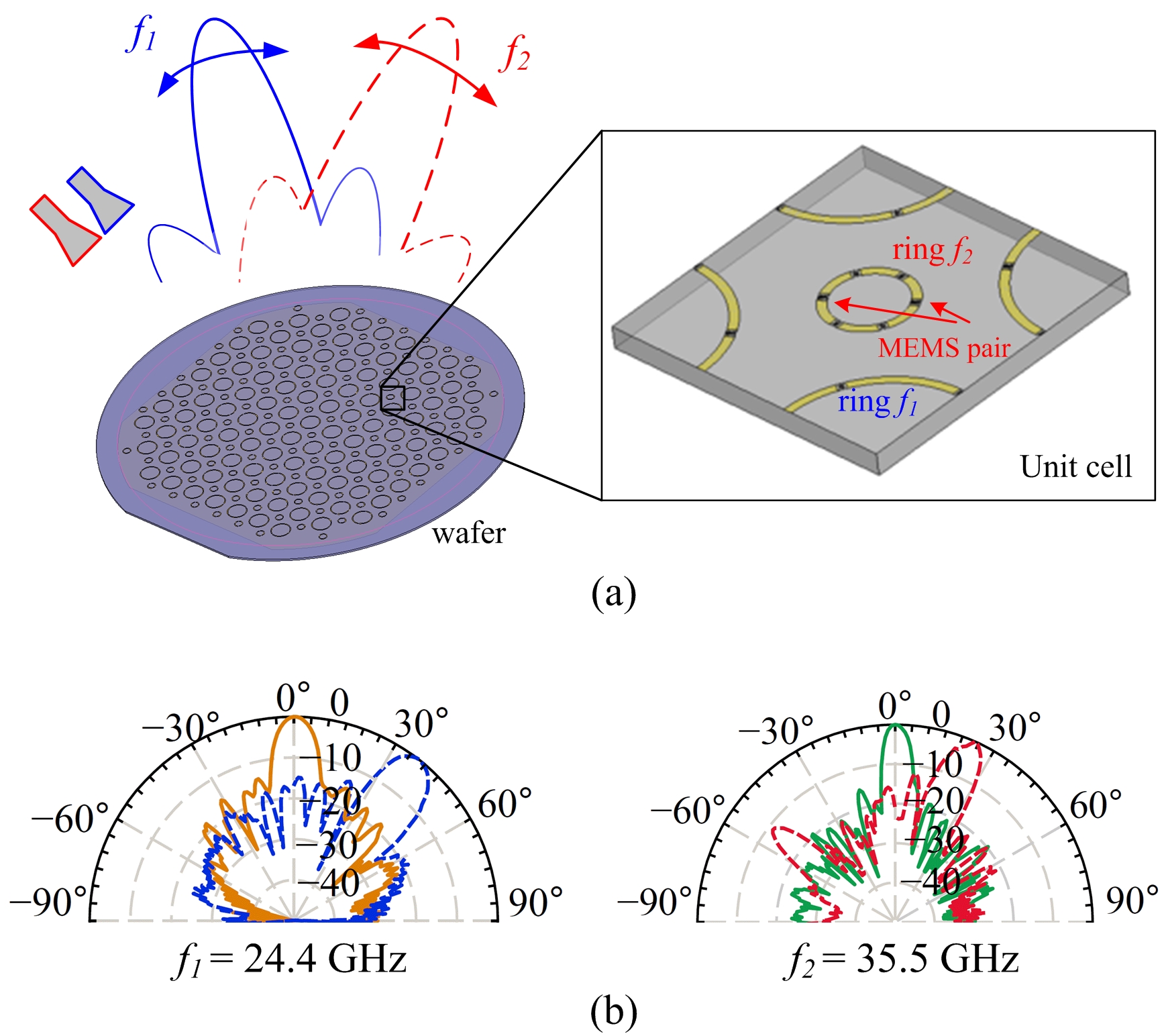}
  \caption{Principle of a reflectarray for independent beam-scanning of two CP beams at different frequencies in the millimeter wave range~\cite{Guc2012a}. (a) scanning is achieved by independently rotating slip-ring elements corresponding to each frequency using MEMS switches (b) measured switched-beams on frozen prototypes at the two operation frequencies.}
  \label{fig:dualband}
\end{figure}

Multi-band reflectarrays have been proposed in the past in fixed configurations. In general such reflectarrays are designed by implementing an ensemble of  reflecting cells for each desired frequency, that are then arranged in a single or over multiple layers depending on the application requirement, in particular on the relative spacing between the desired frequencies. However, multi-band operation in beam-scanning reflectarrays has only been recently considered. A first proof-of-concept of such an operation mode was recently provided considering CP for both frequencies~\cite{Guc2012a}, as depicted in Figure~\ref{fig:dualband}. This was based on the element rotation technique described in Section~\ref{sec:rotation}, using MEMS to electronically implement the required rotation at each cell while preserving low loss at the operating  frequencies 24 GHz and 35 GHz. Measured radiation patterns of frozen full array prototypes are also shown in the figure. 

\subsection{Frequency-agile Reflectarray Elements}

It is well known the performance of reflectarrays in terms of bandwidth is limited, and techniques for wideband operation are more difficult to implement in beam-scanning cells than in fixed array.  In this context, achieving the bandwidths required for some applications might be very challenging.  An example is satellite broadcasting, with downlink / uplink bands of 10.7--12.75 GHz / 14.0--14.5 GHz at \emph{Ku} band.

\begin{figure}[htbp]
  \centering
  \includegraphics[width=\columnwidth]{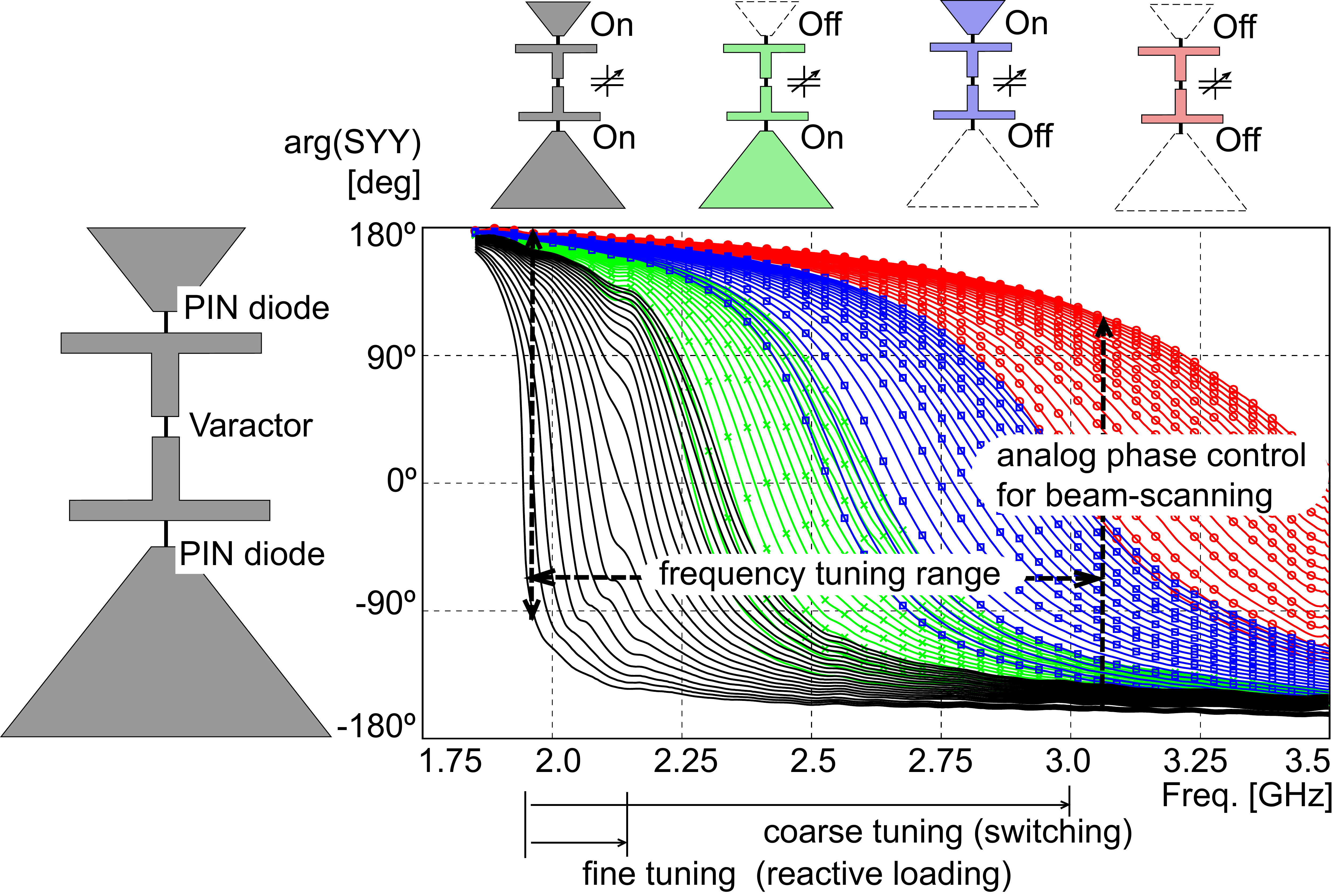}
  \caption{Measured reflection phase of frequency-reconfigurable reflectarray cell in a RWG simulator and operation principle~\cite{Rod2013a}.}
  \label{fig:freqagile}
\end{figure}

Though the bandwidth constraint cannot be overcome by frequency tuning if a very large instantaneous bandwidth is required, frequency reconfiguration is a viable option for selectively receiving / transmitting, or for frequency-hopping systems and cognitive radio. For such a design to be useful, obviously the tuning frequency range must be much wider than the bandwidth achievable with a single-frequency design, depending on the requirements and implementation. In this context a reflectarray cell able to dynamically control the reflection phase at a variable frequency was recently presented in~\cite{Rod2013a}. As shown in the measured results of Figure~\ref{fig:freqagile}, it achieves a continuous tuning range of more than $270^\circ$ of phase range for any desired frequency within a range larger than 1:1.5. The principle of operation of the cell is also symbolically explained in Figure~\ref{fig:freqagile}. The reconfigurable cell combines two switches and a varactor to tune the cell frequency response in a coarse and fine manner, respectively. As a result, the cell can adjust the reflection phase at a variable operating frequency over large and continuous phase-frequency ranges. The length of the cell sections are designed so the spacing between the resonances of the four switch configuration is uniform and identical to the maximum frequency shift induced by the varactor.

\subsection{Active Reflectarrays}
\label{sec:active}

Recently, there has been significant interest in integrating active devices in the form of amplifiers with antenna arrays for a variety of reasons, such as increasing the overall gain of the antenna, compensating for losses, and in the case of transmitters, power-combining for high EIRPs.  At high frequencies, especially in the millimeter-wave frequency range, transistor sizes become very small necessitating the use of power-combining networks to achieve high output powers from power amplifiers.  Losses in transmission line-based power combining networks become pronounced in this frequency range, which spurred considerable interest in spatial power combiners (SPCs) in the 1990s and 2000s~\cite{navarro1996book}\cite{popovic1998}.  %Spatial power combiners essentially consist of an spatially-fed active antenna array with power coupled into and out of the array by feed horns, similar to the situation shown in Figure~\ref{fig:arraylens} but with an additional horn on the output side.  
Essentially, spatial power combiners operate similar to array lenses, except that the output is collimated to be collected by the feed horn on the output side of the lens.

%A subtle variation in the SPC concept is to collimate the output into a far-field pencil beam, rather than for collection by another feed.  This can be achieved in both reflectarray and array lens architectures.  As is the case with all active antennas, stability becomes one of the primary concerns of the design, and the design of the aperture requires special considerations in order to avoid oscillations.

\begin{figure}[hbp]
  \centering
  \subfigure[Co-polarized]
  {\raisebox{4mm}{\includegraphics[scale=0.75]{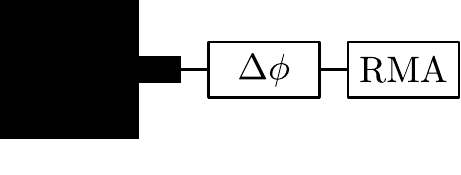}}}
  \hspace{1cm}
  \subfigure[Cross-polarized]
  {\includegraphics[scale=0.75]{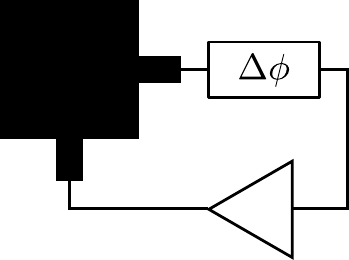}}
  \caption{Active reflectarray unit cell types}
  \label{fig:amptypes}
\end{figure}

In a reflectarray, the active device is engineered into the unit cell such that the reflection coefficient from the unit cell is greater than unity.  There are two ways to achieve this, as illustrated in Figure~\ref{fig:amptypes}.  In a co-polarized reflectarray, the input and output polarizations are the same, necessitating the use of a reflection-mode amplifier (RMA).  Achieving stability in such designs is extremely challenging, since the stability condition is
\begin{equation}
  \label{eq:rma}
  |\Gamma_A(\omega) G_A(\omega)| \le 1
\end{equation}
where $\Gamma_A$ is the input reflection coefficient of the antenna composing the reflectarray unit cell, and $G_A$ is the gain of the reflection-mode amplifier.  This condition must be met over the entire operating frequency range of the amplifier, which can be very challenging.  A more common approach is to utilize a cross-polarized reflectarray design where the input and output polarizations are orthogonal, which employs a two-port dual-polarization antenna as the reflectarray element.  This affords some isolation between the input and isolation, and the stability condition can be approximated as~\cite{kishor2012}
\begin{equation}
  \label{eq:cross}
  |G_A(\omega) S_{12}(\omega)| \le 1
\end{equation}
where $G_A$ is the gain of a two-port amplifier connecting the input and output ports of the antenna, and $S_{12}$ is the coupling between the two ports.  This condition is much easier to meet and has been exploited in the design of fixed-pattern active reflectarrays~\cite{bialkowski2002}\cite{clark2003}.

A natural extension to these designs is to incorporate reconfigurability into the beam pattern.  The use of active devices compensates some of the loss that might be incurred by the tuning mechanism while at the same time increasing the gain of the system and providing power combining capabilities for transmitters.  Though this area is still growing, there have been several cross-polarized designs employing two-port phase-shifters in cascade with amplifiers connecting the two polarization ports of patch antennas.  Experimental results employing phase shifters based on IQ modulators~\cite{cabria2009} and reflection-mode phase shifters~\cite{kishor2012} have been recently presented.  Active array lens designs have also been proposed~\cite{ruggerini2010}, though presently beam-forming is achieved by employing multiple feeds.  %However, this topology most closely resembles a spatial power combiner, and the array lens architecture itself has attracted considerable interested in recent years.  The array lens is the topic of the next section.

\section{A Related Architecture: The Array Lens}
\label{sec:arraylens}

As discussed in Section~\ref{sec:alback}, the array lens topology is a variant of a spatially-fed antenna array whereby one side of the array is illuminated by the feed and the radiation is produced on the opposite side.  Reconfigurable versions of this topology have several advantages over their reflectarray equivalents.  First, array lens designs are free from feed blockage effects, which may be a consideration in small apertures.  Also, in addition to far-field beam-forming, array lenses also have the capacity to form focal points in the vicinity of the lens aperture which can be useful in applications requiring adaptive focusing, such as microwave hyperthermia.

As is the case with electronically tunable reflectarrays, reconfigurable array lens designs can be grouped into similar categories according to the mechanism by which phase shifting is achieved by the unit cell.  Each of these approaches is described in more detail in the following sections.

\subsubsection{Tunable Scatterer Approach}

As described earlier, a key difference between the unit cells in reflectarrays and array lenses is that in an array lens, the phase of the wave must be manipulated with a minimum of reflection and insertion loss, whereas in reflectarrays a strong reflection is generally guaranteed because of the use of a ground plane.  Intrinsic to this process is also the fact that the wave interacts with the scatterer twice during its transit from the feed to the aperture plane, meaning that a single-pole resonator is all that is required to produce nearly $360^\circ$ of phase shift in a reflectarray.  This contrasts significantly with the situation in an array lens.  Resonators the incoming wave interacts with can be seen as introducing a single pole response into the transfer function modelling the input/output characteristic of a unit cell, as is well known in the field of frequency selective surfaces.  Hence, multi-pole designs have been widely employed to tailor the response of FSSs using resonators of different types~\cite{abbaspour-tamijani2004}, or by coupling layers of inductive and capacitive elements~\cite{al-joumayly2010}, to achieve a desired magnitude response for filtering applications.  However, the adaptation of resonators to tunable surfaces has several important implications on the design of RAL unit cells.

Considering the pole/zero behavior in the complex plane is highly useful in understanding the design of array lens elements based on tunable resonators~\cite{lau2011}.  Assuming a resonator pole can be arbitrarily manipulated, the insertion loss and phase are dictated by the distance from, and the angle made with, the pole to the operating frequency point in the complex frequency plane, respectively.  At a fixed operating frequency, in order for the insertion magnitude of the unit cell to remain constant, the pole must be manipulated such that it moves in a circular arc in the left-hand plane around the center frequency.  Achieving such an ideal trajectory is impossible in most designs.  Furthermore, this discussion illustrates that a single pole is capable of contributing, at most, up to $180^\circ$ of phase shift in the transfer function.  Early tunable array lens designs employing only a single-pole response hence achieved very low phase ranges~\cite{torre2008}.  Hence a minimum of two, and preferably three or more, resonators are required to meet the phase requirements of beam-forming.  As a result, even fixed array designs tend to require multi-layer structures of resonators~\cite{ryan2010}, unless one settles for 1-bit ($0^\circ$ / $180^\circ$) phase-shifting which can simplify the cell somewhat~\cite{clemente2012}.

In a similar way, in RALs, designs achieving the required levels of phase agility have been accomplished using different types of resonators.  In many cases, identical resonator elements are desired, because it simplifies the biasing control of the array lens immensely.  In this case, resonators generally need to be separated by a significant electrical distance (e.g.~one quarter-wavelength) in order for the resonators to produce the required phase range while maintaining an acceptable reflection coefficient seen looking into the unit cell~\cite{iturri-hinojosa2010}--\nocite{boccia2011}\nocite{lau2012b}\cite{jiang2012}.  However, the increased electrical distance not only increases the physical thickness of the lens, it also introduces an inter-layer coupling mechanism which has been shown to potentially lead to spurious radiation in undesired directions~\cite{lau2012b}.  

The other option is to use an arrangement of dissimilar resonators in order to alleviate the need to separate the resonators by a large distance.  For example, tunable patch resonators can be coupled to a capacitively-tuned slot resonator to effectively realized a triple-pole response using a very thin structure~\cite{lau2011}.  Theoretically, achieving thin tunable array lenses is possible by closely coupling together capacitive and inductive surfaces to form a tunable FSS~\cite{al-joumayly2010}, but would require tunable capacitances on the capacitive surface (straightforward to implement) and tunable inductors (more challenging to implement) on the inductive surface to achieve best overall performance.  %The latter is generally requires exotic devices (such as tunable MEMS inductors) or equivalent implementations based on tunable capacitances that the resulting bandwidth would be sub-optimal.  In any case, 
The main disadvantage of an approach using dissimilar layers the layers need to be tuned separately each other in order to form the right pole trajectories to maximize the phase range while minimizing the insertion loss through the unit cell.  This can complicate the biasing and control of such surfaces, and combined with the need for thin array lenses, has motivated research on the next approach.

%\textbf{fluidic?}

\subsubsection{Guided-Wave Approach}

In this approach, the array elements composing the input of the array lens are connected to the array elements composing the output of the array lens via a two-port guided-wave network.  %Traditionally, in the first array lenses, this network was simply a delay line whose length was appropriate chosen to achieve the collimation of the feed required on the output side of the lens.  However, 
In reconfigurable designs, this network must be electronically tunable, and, as we have seen in Section~\ref{sec:active}, can potentially incorporate gain as well.

Only a handful of reconfigurable array lenses of this type have been experimentally demonstrated.  Borrowing the terminology of SPCs, a ``tray'' approach can be taken whereby phase-shifting circuits are integrated with the input and output faces of the lens in a three-dimensional manner~\cite{torre2010}, with the primary drawback being a thick structure that is more difficult to manufacture.  Other approaches tend towards ``tile'' forms of integration, employing designs that use varactor diode-tuned bridged-T phase shifters~\cite{lau2012b}\cite{lau2012a} or MEMS switches to adjust the delay through a bandpass structure~\cite{cheng2009}.  

\begin{figure}[htbp]
  \centering
  %\subfigure[External view]
  %{\includegraphics[width=0.45\columnwidth]{jonlens1}}
  %\subfigure[Internal view]
  %{\includegraphics[width=0.45\columnwidth]{jonlens2}}
  \includegraphics[width=\columnwidth]{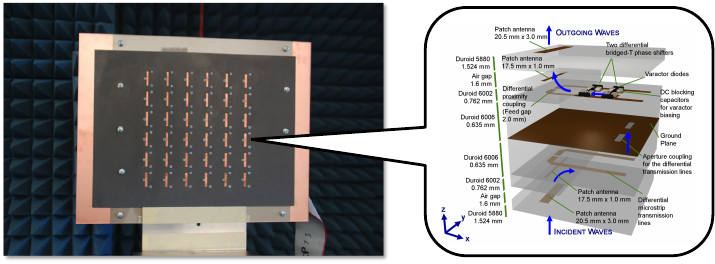}
  \caption{Experimental array lens~\cite{lau2012b}}
  \label{fig:jonlens}
\end{figure}

While research on tunable array lenses is still in its infancy, the potentially thin nature and bandwidth of the guided-wave approach makes it an attractive topology.  Figure~\ref{fig:jonlens} shows a recent example of an experimental prototype exhibiting a 10\% fractional bandwidth at 5~GHz.

\subsubsection{Element Rotation Approach}

Reconfigurable array lenses for handling CP waves are still at an early developmental stage.   Fixed CP array lenses have been explored that incur the phase shift by manipulating the LP components of a CP wave~\cite{phillion2011} or by employing the element rotation technique~\cite{okane2009}--\nocite{euler2010}\cite{kaouach2011}.  However, to our knowledge, array lenses employing this approach have yet to be realized in reconfigurable form.

\section{Ongoing and Future Challenges}
\label{sec:ongoing}

\subsection{Bandwidth extension and transformation optics approaches}
\label{sec:bandwidth}

A well-known limitation of reflectarrays and array lenses is their limited operating bandwidth, and this is currently a very active area of research.  Bandwidth limitations fundamentally originate from the fact that in order to achieve ideal bandwidth characteristics, the array elements must produce a true time delay (TTD) response.  Most reflectarray and array lens elements can only approximate such a response over a narrow bandwidth.  Addressing this limitation for RRAs and RALs is particularly challenging, for reasons that are outlined below.

In the case of reflectarrays, bandwidth constraints are usually alleviated by employing one of two approaches.  The first is to increase the phase bandwidth of the elements by attempting to approximate the TTD response over a limited band. This can be achieved by employing multi-resonant elements, such as stacked patches~\cite{encinar2003} and concentric loops~\cite{chaharmir2010} to name a few popular techniques.  These approaches have been applied with success in fixed reflectarrays, improving bandwidth from around 3\% in the case of single-resonant elements to around 12\% in the case of multi-resonant elements.  The challenge with adapting this to reconfigurable designs is that these element designs employed coupled resonators to improve the bandwidth of the element.  By varying the size and shape of the individual resonators, one not only changes the resonant frequency of each constituent resonator, but also the inter-resonator coupling.  In electronically tunable variants, the resonator frequency can be readily controlled through integration with tunable components, as we have seen in Section~\ref{sec:resonator}, but the inter-resonator electromagnetic coupling is not affected significantly by the tuning because the geometry of the resonator remains fixed.  Hence, more sophisticated tuning techniques, that also employing tunable devices to vary the inter-resonator coupling, are required to improve the bandwidth of multi-resonant element designs~\cite{liu2010}.  Using this technique, the element phase bandwidth can be effectively multiplied by the number of resonators employed in the unit cell.

Resonators composing the element can also be designed to be co-planar, which tends to reduce the coupling effect and allow the individual resonances of each element to play a larger role in controlling the bandwidth.  Unit cells composed of three parallel dipole resonators situated over a tunable liquid crystal substrate have been proposed~\cite{perez-palomino2012} and experimentally verified~\cite{perez-palomino2013} as a means for extending the bandwidth of LC cells to 8\%.

The desired TTD behavior can be easily achieved using the guided-wave approach.  Fixed designs employing aperture-coupled microstrip delay lines~\cite{carrasco2007} have been shown to substantially improve the bandwidth of reflectarrays to the 10\% range.  As discussed earlier, this approach can be adapted to provide beam-steering~\cite{carrasco2011}.  However, since the amount of time delay that can be created within the space constraints of the cell is limited, and the bandwidth of such element is also conditional to the  wideband matching of the resonating element with the phase shifter.

For array lenses, the bandwidth similarly depends on their implementation.  Classical array lenses were based on transmission lines connecting the input and output elements.  However, array lenses based on resonant FSS structures have smaller bandwidths, though this situation has been alleviated through the use of miniaturized element FSSs (MEFSSs)~\cite{al-joumayly2010}.  

Similar to reflectarrays, wideband RALs must employ either wideband phase shifters coupled to wideband elements, or reconfigurable TTD structures.  As wideband element and phase shifter designs are widespread in the literature, ultimately the bandwidth limitations stem from issues arising in the compact integration of the elements with the phase shifters in the former approach~\cite{lau2012b}.  Regarding the latter approach, again, switched structures, which trade off bit resolution for bandwidth, attempt to implement TTD structures for enhanced bandwidth~\cite{cheng2009}.

\begin{figure}[htbp]
  \centering
  \subfigure[Virtual space]
  {\raisebox{2mm}{\includegraphics[scale=0.23]{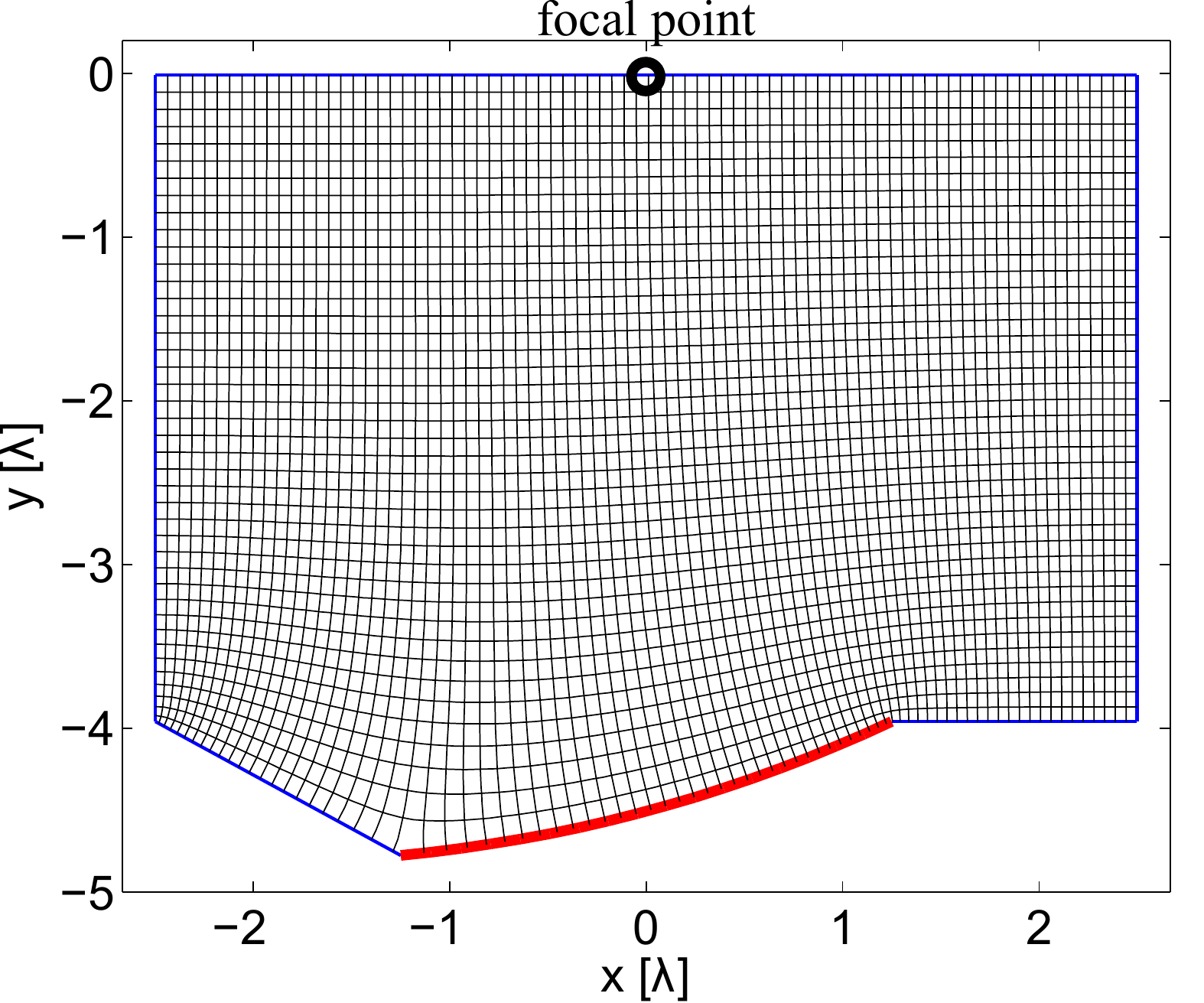}}}
  \subfigure[Physical space]
  {\includegraphics[scale=0.23]{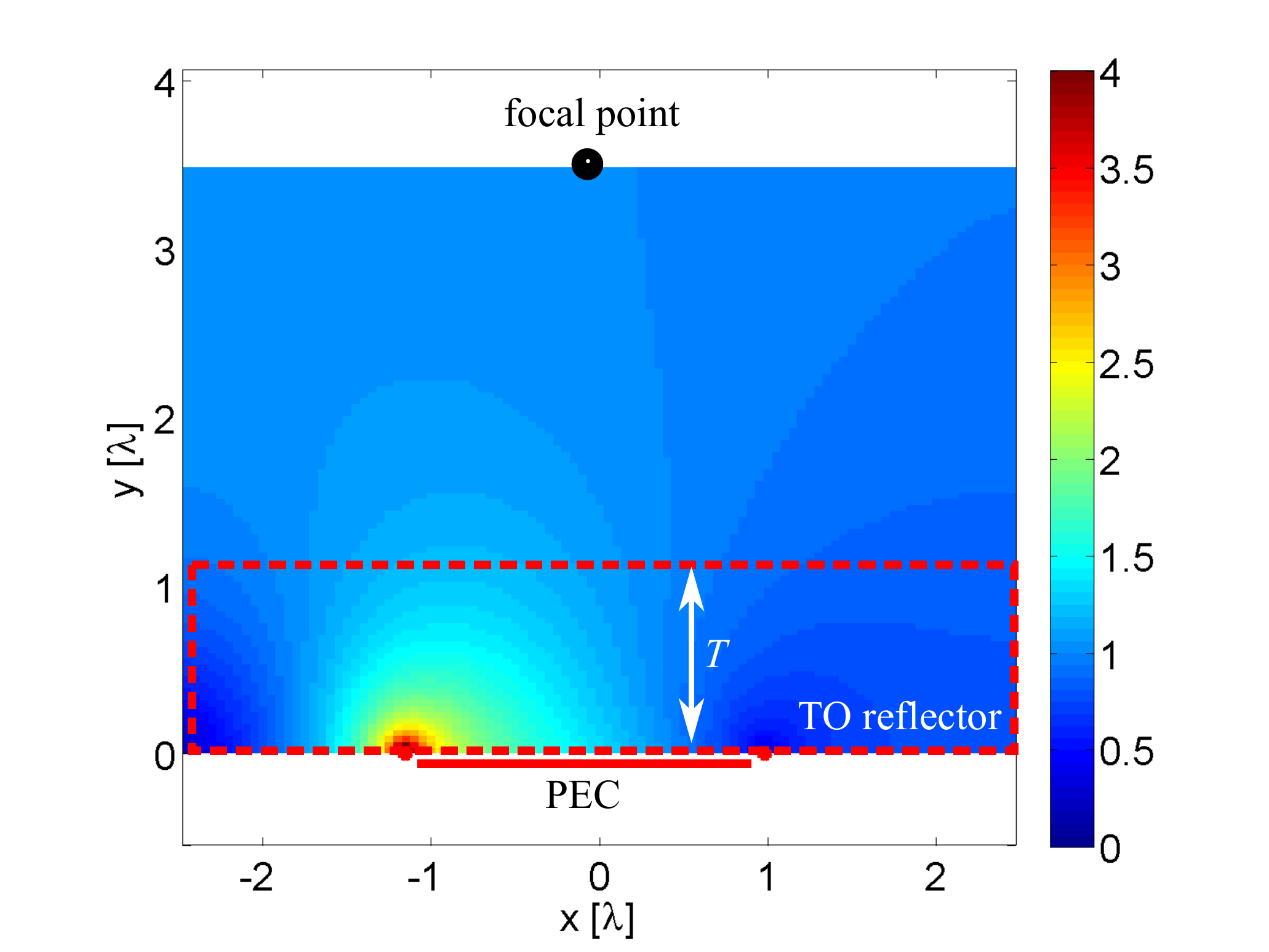}}
  \caption{Transformation between virtual (a) and physical spaces (b) in the TO approach}
  \label{fig:to}
\end{figure}

Ways to further improve the bandwidth of spatially-fed apertures is an area of active research, with particularly promising solutions arising from the field of transformation optics (TO)~\cite{kundtz2011}.  In the synthesis of wideband apertures using a TO approach, the desired field transformation (e.g.~from a spherical feed to a plane-wave pencil-beam) is defined spatially in one spatial domain, and the metric-invariant property of Maxwell's equations can be employed to achieve the same wave propagation in another spatial domain filled with a region of inhomogeneous dielectrics.  In the case of reflectarrays, this region is a cover that is placed over a flat reflector (ground plane) composing the reflector, while in array lenses, the region defines the lens itself.  Conceptually, the case in Figure~\ref{fig:to} illustrates the transformation between the warped space (virtual space) and the regular space (physical space) succinctly.  Essentially, the material cover implements a spatially-distributed TTD system to ensure all rays from the feed are delayed appropriately by the aperture.  Moreover, since the transformation does not yield large zones of materials exhibiting unusual electromagnetic properties (e.g.~a refractive index less than 1), the dispersion associated with such materials is avoided yielding a potentially very broadband transformation device.  This approach has been used to successfully design flat reflectors~\cite{tang2010} and lenses~\cite{yang2011}\cite{wu2012} in principle.  

Researchers have contemplated beam-steerable versions~\cite{yang2011awpl} provided the electromagnetic properties of the dielectric region can be manipulated.  Effective material implementations based on metallic conductors have recently been investigated as a means for realizing the dielectric region, while providing reflectors with very large potential beam-scanning ranges~\cite{liang2013oe}.  A similar implementation can be pursued for array lenses.  Such conductors can be loaded with tunable components to yield ultra-wideband implementations of spatially-fed apertures, which may benefit many applications in the future.  A disadvantage of this approach is that the dielectric region tends to be quite thick in order to facilitate the field transformation, creating interesting challenges for future research.

\subsection{Mitigating Nonlinear Behavior}

A major challenge facing designers of electronically tunable reflectarrays and array lenses is the linearity of the underlying tuning technology.  Table~\ref{tab:technologies} summarizes the technologies and shows their corresponding linearity.  RRAs and RALs are being proposed for satellite and RADAR applications, which employ transmitters with very high output powers.  Hence, the illumination of the aperture may induce nonlinear behaviour in the underlying technologies, leading to harmonic and inter-modulation distortion (IMD).  In the area of communication systems, particularly satellite communications, there are strict limitations on the allowable adjacent channel interference and harmonic levels produced by the transmitter.  Even passive inter-modulation is a source of compliance failure in these transmitters, indicating that ultra-linear tuning technologies must be employed should reconfigurable apertures find practical use in these applications.

Semiconductor technologies, while mature and widespread, are the most prone to this problem.  Varactor diodes integrated in the aperture, for example, can easily have their capacitance modulated at the frequency of the illuminating signal, causing phase modulation that manifests itself in terms of distortion of the scattered signal.  For example, the IMD performance of a single-pole reflectarray element has been evaluated in a waveguide simulator~\cite{hum2007}, illustrating that even under modest illumination power levels, the creation of odd-order distortion is significant for varactor diode-tuned elements which ultimately may relegate such apertures to receive-only applications.  Researchers are keenly aware of this and efforts to document the linearity of new designs is undertaken in most recent studies~\cite{clemente2012}.

Nevertheless, addressing this challenge remains a fundamental concern for designers.  Ultimately, MEMS technology may be the solution to this problem~\cite{hum2006}\cite{sorrentino2009}.  Other exotic materials possessing large relaxation times, such as liquid crystals, may also be contenders as the underlying technologies mature.

\subsection{Very large apertures realized using compound apertures}

Realizing very large, high-gain apertures composed of reconfigurable unit cells, while theoretically possible, poses many practical constraints in terms of the number of devices required (and associated cost), bias network complexities, and in some cases, device power requirements.  However, many high-gain cost-effective solutions can be realized by combining adaptive spatially-fed arrays with fixed apertures.  The most common of these is a parabolic or flat main reflector (which itself could be realized as a fixed reflectarray) illuminated by a sub-reflector composed of a reflectarray.  Such dual-reflector combinations can be used to emulate their traditional counterparts, such as the Cassegrain antennas~\cite{rajagopalan2012} and offset reflectors~\cite{tienda2011}.  In particular, employing a reconfigurable sub-reflectarray (or array lens) is an effective way to manage the cost of the antenna system, since the primary large-area aperture is not reconfigurable.  The tradeoff is that the overall scanning range of the antenna system is reduced, often to just a few degrees depending on the system geometry.  However, many applications do not require large scan ranges, such as atmospheric limb-sounding, certain satellite applications, etc.  Liquid-crystal reflectarrays have successfully been integrated as sub-reflectors into dual-reflector systems for such purposes~\cite{hu2009}, and extensions of this technique to other configurations may be a practical way forward in realizing very high gain reconfigurable apertures in the future.

\subsection{Towards Terahertz and Optical Frequencies}

The application of the reflectarray and lens-array concepts to higher frequencies has recently attracted significant attention. For instance, at optical frequencies fixed configurations have been proposed using metals in the plasmonic regime~\cite{farmahini-farahani2013}--\nocite{Niu2013a}\nocite{Xia2012a}\cite{Yu2011a} and lower-loss dielectric scatterers~\cite{Lon2013a}. Though higher operation frequencies obviously entail significant novel practical considerations, the operating principle is essentially the same as in prior art at lower frequencies.  

Dynamic beam control is now also considered at higher frequencies. Potential applications at terahertz frequencies are numerous both for sensing and communication, where the reflectarray and lens-array concept should provide a low-loss and relatively simple solution for electronic beam control. At optical frequencies, applications mostly relate to sensing but flexible free-space interconnects or even visible light communication could be of interest in the future.

Early attempts to beam-scanning at particularly high frequencies include the on-going effort to produce a MEMS-based reflectarray at 120 GHz~\cite{Tam2012a}.  However, such frequencies can be considered as an upper bound for the applicability of standard RF-MEMS technology, among others, because the MEMS elements become too electrical large to efficiently be integrated within the array unit cell.  Therefore, it is particularly relevant to consider enabling technologies based on reconfigurable materials. For instance, there have been experimental demonstrations of reflectarray unit cells using LC crystals, based on the principle that the LC anisotropic permittivity tensor is modified by an applied bias field~\cite{perez-palomino2013}. Though this demonstration was done at sub-millimeter-wave frequencies, the use of LC has proven very efficient at optical frequencies in display applications and thus should also be applicable to the reflectarray concept. Second, the use of graphene for beam-scanning at 1.3 THz has been proposed~\cite{Car2013b}. In this case, graphene's 2D complex surface impedance is dynamically controlled by applying a bias voltage to a nearby electrode (so-called graphene `field effect') to achieve dynamic phase control. These LC and graphene cell concepts are based on the typical resonant cell topology consisting of a conductive patch resonator above a substrate, with the following notable difference. In the case of LC it is the substrate parameters that are controlled, while in the case of graphene the substrate is fixed but the resonance is altered by the change in the complex conductivity of the graphene patch. 

These results are encouraging but preliminary in terms of the highest frequency achievable and experimental implementations. Moreover, other enabling technologies should be studied and compared. Therefore the implementation of beam-scanning RRAs and RALs at terahertz and optical frequencies constitute an important and exciting playground, where technological issues are bound to play an extremely important role compared to lower frequency applications.

%\textbf{What about \cite{ginn2007},\cite{yang2012}?}

\section{Conclusions}
\label{sec:conclusions}

Recent progress on reconfigurable reflectarrays and array lenses is enabling these architectures to compete with established antenna beam-forming technologies such as phased arrays.  Essentially, these architectures combine the best features of arrays and aperture antennas, yielding an efficient yet cost-effective platform for high-gain adaptive beam-forming and beam-synthesis.  This article has reviewed the key technologies and approaches for realizing these antennas, and identified some key capabilities of RRAs and RALs provide that are unique to these antenna types and not easily replicated using other platforms.  

Looking forward, this research area is full of exciting future research possibilities.  This paper has outlined some of the present shortcomings of the technologies in terms of bandwidth, operating frequency, hardware cost, and linearity, and continued exploration of the underlying technologies and designs of RRAs and RALs will make them even more promising candidates for host of applications in the future.

%\bibliographystyle{IEEEtran}
%\bibliography{IEEEabrv,references}

\begin{thebibliography}{100}
\providecommand{\url}[1]{#1}
\csname url@samestyle\endcsname
\providecommand{\newblock}{\relax}
\providecommand{\bibinfo}[2]{#2}
\providecommand{\BIBentrySTDinterwordspacing}{\spaceskip=0pt\relax}
\providecommand{\BIBentryALTinterwordstretchfactor}{4}
\providecommand{\BIBentryALTinterwordspacing}{\spaceskip=\fontdimen2\font plus
\BIBentryALTinterwordstretchfactor\fontdimen3\font minus
  \fontdimen4\font\relax}
\providecommand{\BIBforeignlanguage}[2]{{%
\expandafter\ifx\csname l@#1\endcsname\relax
\typeout{** WARNING: IEEEtran.bst: No hyphenation pattern has been}%
\typeout{** loaded for the language `#1'. Using the pattern for}%
\typeout{** the default language instead.}%
\else
\language=\csname l@#1\endcsname
\fi
#2}}
\providecommand{\BIBdecl}{\relax}
\BIBdecl

\bibitem{berry1963}
D.~G. Berry, R.~G. Malech, and W.~A. Kennedy, ``The reflectarray antenna,''
  \emph{{IEEE} Trans. Antennas Propag.}, vol.~11, no.~6, pp. 645--651, Nov.
  1963.

\bibitem{huang1991}
J.~Huang, ``Microstrip reflectarray,'' in \emph{1991 Antennas Propag. Soc. Int.
  Symp. Dig.}, vol.~2, Jun. 1991, pp. 612--615.

\bibitem{huang2007book}
J.~Huang and J.~A. Encinar, \emph{Reflectarray Antennas}.\hskip 1em plus 0.5em
  minus 0.4em\relax Wiley - IEEE Press, 2007.

\bibitem{pozar1997}
D.~M. Pozar, S.~D. Targonski, and H.~D. Syrigos, ``Design of millimeter wave
  microstrip reflectarrays,'' \emph{{IEEE} Trans. Antennas Propag.}, vol.~45,
  no.~2, pp. 287--296, Feb. 1997.

\bibitem{chaharmir2003}
M.~R. Chaharmir, J.~Shaker, M.~Cuhaci, and A.~Sebak, ``Reflectarray with
  variable slots on ground plane,'' \emph{IEE Proc. Microwaves Antennas
  Propag.}, vol. 150, no.~6, pp. 436--439, Dec 2003.

\bibitem{bialkowski2008}
M.~Bialkowski and K.~H. Sayidmarie, ``Investigations into phase characteristics
  of a single-layer reflectarray employing patch or ring elements of variable
  size,'' \emph{{IEEE} Trans. Antennas Propag.}, vol.~56, no.~11, pp.
  3366--3372, 2008.

\bibitem{bozzi2003}
M.~Bozzi, S.~Germani, and L.~Perregrini, ``Performance comparison of different
  element shapes used in printed reflectarrays,'' \emph{{IEEE} Antennas
  Wireless Propag. Lett.}, vol.~2, no.~1, pp. 219--222, 2003.

\bibitem{chang1995}
D.~C. Chang and M.~C. Huang, ``Multiple-polarization microstrip reflectarray
  antenna with high efficiency and low cross-polarization,'' \emph{{IEEE}
  Trans. Antennas Propag.}, vol.~43, no.~8, pp. 829--834, Aug. 1995.

\bibitem{huang1998}
J.~Huang and R.~J. Pogorzelski, ``A {K}a-band microstrip reflectarray with
  elements having variable rotation angles,'' \emph{{IEEE} Trans. Antennas
  Propag.}, vol.~46, no.~5, pp. 650--656, May 1998.

\bibitem{martynyuk2004}
A.~Martynyuk, J.~Martinez~Lopez, and N.~Martynyuk, ``Spiraphase-type
  reflectarrays based on loaded ring slot resonators,'' \emph{{IEEE} Trans.
  Antennas Propag.}, vol.~52, no.~1, pp. 142--153, 2004.

\bibitem{encinar2001}
J.~Encinar, ``Design of two-layer printed reflectarrays using patches of
  variable size,'' \emph{{IEEE} Trans. Antennas Propag.}, vol.~49, no.~10, pp.
  1403--1410, Oct. 2001.

\bibitem{encinar2003}
J.~A. Encinar and J.~A. Zornoza, ``Broadband design of three-layer printed
  reflectarrays,'' \emph{{IEEE} Trans. Antennas Propag.}, vol.~51, no.~7, pp.
  1662--1664, Jul. 2003.

\bibitem{chaharmir2010}
M.~Chaharmir, J.~Shaker, N.~Gagnon, and D.~Lee, ``Design of broadband, single
  layer dual-band large reflectarray using multi open loop elements,''
  \emph{{IEEE} Trans. Antennas Propag.}, vol.~58, no.~9, pp. 2875--2883, 2010.

\bibitem{carrasco2008}
E.~Carrasco, J.~Encinar, and M.~Barba, ``Bandwidth improvement in large
  reflectarrays by using true-time delay,'' \emph{{IEEE} Trans. Antennas
  Propag.}, vol.~56, no.~8, pp. 2496--2503, 2008.

\bibitem{hsu2007}
S.-H. Hsu, C.~Han, J.~Huang, and K.~Chang, ``An offset linear-array-fed ku/ka
  dual-band reflectarray for planet cloud/precipitation radar,'' \emph{{IEEE}
  Trans. Antennas Propag.}, vol.~55, no.~11, pp. 3114--3122, 2007.

\bibitem{yu2010}
A.~Yu, F.~Yang, A.~Elsherbeni, and J.~Huang, ``Experimental demonstration of a
  single layer tri-band circularly polarized reflectarray,'' in \emph{Proc.
  2010 IEEE Antennas Propag. Soc. Int. Symp.}, 2010.

\bibitem{pozar2007}
D.~Pozar, ``Wideband reflectarrays using artificial impedance surfaces,''
  \emph{Electron. Lett.}, vol.~43, no.~3, pp. 148--149, 2007.

\bibitem{sievenpiper1999}
D.~Sievenpiper, L.~Zhang, R.~F. Broas, N.~G. Alexopolous, and E.~Yablonovitch,
  ``High-impedance electromagnetic surfaces with a forbidden frequency band,''
  \emph{{IEEE} Trans. Microw. Theory Tech.}, vol.~47, no.~11, pp. 2059--2074,
  Nov. 1999.

\bibitem{nayeri2010awpl}
P.~Nayeri, F.~Yang, and A.~Elsherbeni, ``Broadband reflectarray antennas using
  double-layer subwavelength patch elements,'' \emph{{IEEE} Antennas Wireless
  Propag. Lett.}, vol.~9, pp. 1139--1142, 2010.

\bibitem{edalati2012}
A.~Edalati and K.~Sarabandi, ``Wideband reflectarray antenna based on
  miniaturized element frequency selective surfaces,'' in \emph{Proc. 2012
  Europ. Conf. Antennas Propag. (EuCAP2012)}, 2012, pp. 362--364.

\bibitem{kock1949}
W.~E. Kock, ``Path-length microwave lenses,'' \emph{Proc. IRE}, vol.~37, no.~8,
  pp. 852--855, 1949.

\bibitem{mcgrath1986}
D.~McGrath, ``Planar three-dimensional constrained lenses,'' \emph{{IEEE}
  Trans. Antennas Propag.}, vol.~34, no.~1, pp. 46--50, 1986.

\bibitem{pozar1996}
D.~M. Pozar, ``Flat lens antenna concept using aperture coupled microstrip
  patches,'' \emph{Electron. Lett.}, vol.~32, no.~23, pp. 2109--2111, 1996.

\bibitem{popovic1998}
Z.~Popovic and A.~Mortazawi, ``Quasi-optical transmit/receive front ends,''
  \emph{{IEEE} Trans. Microw. Theory Tech.}, vol.~46, no.~11, pp. 1964--1975,
  1998.

\bibitem{barba2006}
M.~Barba, E.~Carrasco, and J.~A. Encinar, ``Suitable planar transmit-arrays in
  {X}-band,'' in \emph{Proc. 2006 Europ. Conf. Antennas Propag. (EuCAP2006)},
  2006.

\bibitem{abbaspour-tamijani2004}
A.~Abbaspour-Tamijani, K.~Sarabandi, and G.~M. Rebeiz, ``Antenna-filter-antenna
  arrays as a class of bandpass frequency-selective surfaces,'' \emph{{IEEE}
  Trans. Microw. Theory Tech.}, vol.~52, no.~8, pp. 1781--1789, 2004.

\bibitem{phillion2011}
R.~Phillion and M.~Okoniewski, ``Lenses for circular polarization using planar
  arrays of rotated passive elements,'' \emph{{IEEE} Trans. Antennas Propag.},
  vol.~59, no.~4, pp. 1217 --1227, 2011.

\bibitem{milne1980}
R.~Milne, ``Dipole array lens antenna,'' in \emph{Antennas Propag. Soc. Int.
  Symp.}, vol.~18, 1980, pp. 576--579.

\bibitem{ryan2010}
C.~G.~M. Ryan, M.~R. Chaharmir, J.~Shaker, J.~R. Bray, Y.~M.~M. Antar, and
  A.~Ittipiboon, ``A wideband transmitarray using dual-resonant double square
  rings,'' \emph{{IEEE} Trans. Antennas Propag.}, vol.~58, no.~5, pp. 1486
  --1493, 2010.

\bibitem{al-joumayly2011}
M.~Al-Joumayly and N.~Behdad, ``Wideband planar microwave lenses using
  sub-wavelength spatial phase shifters,'' \emph{{IEEE} Trans. Antennas
  Propag.}, vol.~59, no.~12, pp. 4542--4552, 2011.

\bibitem{cooley1997}
M.~E. Cooley, J.~F. Walker, D.~G. Gonzalez, and G.~E. Pollon, ``Novel
  reflectarray element with variable phase characteristics,'' \emph{IEE Proc.
  Microwaves Antennas Propag.}, vol. 144, no.~2, pp. 149--151, May 1997.

\bibitem{chaharmir2006tap}
M.~Chaharmir, J.~Shaker, M.~Cuhaci, and A.~Sebak, ``Novel
  photonically-controlled reflectarray antenna,'' \emph{{IEEE} Trans. Antennas
  Propag.}, vol.~54, no.~4, pp. 1134--1141, 2006.

\bibitem{Leg2009a}
H.~Legay, D.~Bresciani, E.~Girard, R.~Chiniard, E.~Labiole, O.~Vendier, and
  G.~Caille, ``Recent developments on reflectarray antennas at {T}hales
  {A}lenia {S}pace,'' in \emph{Proc. 2009 Europ. Conf. Antennas Propag.
  (EuCAP2009)}, 2009, pp. 2515--2519.

\bibitem{Lon2011a}
S.~Long and G.~Huff, ``A fluidic loading mechanism for phase reconfigurable
  reflectarray elements,'' \emph{{IEEE} Antennas Wireless Propag. Lett.},
  vol.~10, pp. 876--879, 2011.

\bibitem{Car2013a}
E.~Carrasco and J.~Perruisseau-Carrier, ``Reflectarray antenna at terahertz
  using graphene,'' \emph{{IEEE} Antennas Wireless Propag. Lett.}, vol.~12, pp.
  253--256, 2013.

\bibitem{hum2007}
S.~V. Hum, M.~Okoniewski, and R.~J. Davies, ``Modeling and design of
  electronically tunable reflectarrays,'' \emph{{IEEE} Trans. Antennas
  Propag.}, vol.~55, no.~8, pp. 2200--2210, Aug. 2007.

\bibitem{Kam2011a}
H.~Kamoda, T.~Iwasaki, J.~Tsumochi, T.~Kuki, and O.~Hashimoto, ``60-{GHz}
  electronically reconfigurable large reflectarray using single-bit phase
  shifters,'' \emph{{IEEE} Trans. Antennas Propag.}, vol.~59, no.~7, pp.
  2524--2531, 2011.

\bibitem{Leg2003a}
H.~Legay, B.~Pinte, M.~Charrier, A.~Ziaei, E.~Girard, and R.~Gillard, ``A
  steerable reflectarray antenna with {MEMS} controls,'' in \emph{Proc. 2003
  IEEE Int. Symp. Phased Array Syst. Tech.}, 2003, pp. 494--499.

\bibitem{hum2006}
S.~V. Hum, G.~McFeetors, and M.~Okoniewski, ``Integrated {MEMS} reflectarray
  elements,'' in \emph{Proc. 2006 Europ. Conf. Antennas Propag. (EuCAP2006)},
  Nov. 2006.

\bibitem{perruisseau-carrier2008}
J.~Perruisseau-Carrier and A.~Skriverviky, ``Monolithic {MEMS}-based
  reflectarray cell digitally reconfigurable over a 360$^\circ$ phase range,''
  \emph{{IEEE} Antennas Wireless Propag. Lett.}, vol.~7, pp. 138--141, 2008.

\bibitem{Rom2007a}
R.~Romanofsky, ``Advances in scanning reflectarray antennas based on
  ferroelectric thin-film phase shifters for deep-space communications,''
  \emph{Proc. {IEEE}}, vol.~95, no.~10, pp. 1968--1975, 2007.

\bibitem{Mai2005a}
R.~Mailloux, \emph{Phased array antenna handbook}.\hskip 1em plus 0.5em minus
  0.4em\relax Boston: Artech House, 2005.

\bibitem{Eba2009a}
S.~Ebadi, R.~Gatti, and R.~Sorrentino, ``Linear reflectarray antenna design
  using 1-bit digital phase shifters,'' in \emph{Proc. 2009 Europ. Conf.
  Antennas Propag. (EuCAP2009)}, 2009, pp. 3729--3732.

\bibitem{Deb2013a}
T.~Debogovi\'{c}, S.~Su\v{s}ac, and J.~Perruisseau-Carrier, ``A simple
  reflectarray cell for 1-bit phase control and polarization flexibility,'' in
  \emph{Proc. 2013 Europ. Conf. Antennas Propag. (EuCAP2013)}, Gothenburg,
  Sweden, 2013.

\bibitem{montori2010}
S.~Montori, F.~Cacciamani, R.~Gatti, E.~Carrasco, M.~Barba, J.~Encinar, and
  R.~Sorrentino, ``Wideband dual-polarization reconfigurable elementary cell
  for electronic steerable reflectarray at {K}u-band,'' in \emph{Proc. 2010
  Europ. Conf. Antennas Propag. (EuCAP2010)}, 2010.

\bibitem{perez-palomino2013}
G.~Perez-Palomino, P.~Baine, R.~Dickie, M.~Bain, J.~Encinar, R.~Cahill,
  M.~Barba, and G.~Toso, ``Design and experimental validation of liquid
  crystal-based reconfigurable reflectarray elements with improved bandwidth in
  {F}-band,'' \emph{{IEEE} Trans. Antennas Propag.}, vol.~61, no.~4, pp.
  1704--1713, 2013.

\bibitem{Car2013b}
E.~Carrasco, M.~Tamagnone, and J.~Perruisseau-Carrier, ``Tunable graphene
  reflective cells for thz reflectarrays and generalized law of reflection,''
  \emph{Appl. Phys. Lett.}, vol. 102, no.~10, pp. 104\,103--104\,103--4, 2013.

\bibitem{perruisseau-carrier2010mtt}
J.~Perruisseau-Carrier, F.~Bongard, R.~Golubovic-Niciforovic,
  R.~Torres-S{\'a}nchez, and J.~Mosig, ``Contributions to the modeling and
  design of reconfigurable reflecting cells embedding discrete control
  elements,'' \emph{{IEEE} Trans. Microw. Theory Tech.}, vol.~58, no.~6, pp.
  1621--1628, 2010.

\bibitem{perruisseau-carrier2010eucap}
J.~Perruisseau-Carrier, E.~Girard, and H.~Legay, ``Analysis of a reconfigurable
  reflectarray cell comprising a multitude of {MEMS} control elements,'' in
  \emph{Proc. 2010 Europ. Conf. Antennas Propag. (EuCAP2010)}, 2010.

\bibitem{salti2010}
H.~Salti, E.~Fourn, R.~Gillard, and H.~Legay, ``Minimization of {MEMS}
  breakdowns effects on the radiation of a {MEMS} based reconfigurable
  reflectarray,'' \emph{{IEEE} Trans. Antennas Propag.}, vol.~58, no.~7, pp.
  2281--2287, 2010.

\bibitem{You2012a}
M.~Yousefbeiki and J.~Perruisseau-Carrier, ``A practical technique for
  accurately modeling reconfigurable lumped components in commercial full-wave
  solvers [{EurAAP} corner],'' \emph{{IEEE} Antennas Propag. Mag.}, vol.~54,
  no.~5, pp. 298--303, 2012.

\bibitem{hum2005}
S.~Hum, M.~Okoniewski, and R.~Davies, ``Realizing an electronically tunable
  reflectarray using varactor diode-tuned elements,'' \emph{{IEEE} Microw.
  Wireless Compon. Lett.}, vol.~15, no.~6, pp. 422--424, 2005.

\bibitem{riel2007}
M.~Riel and J.~Laurin, ``Design of an electronically beam scanning reflectarray
  using aperture-coupled elements,'' \emph{{IEEE} Trans. Antennas Propag.},
  vol.~55, no.~5, pp. 1260--1266, 2007.

\bibitem{rodriguez-zamudio2012}
J.~Rodriguez-Zamudio, J.~Martinez-Lopez, J.~Rodriguez-Cuevas, and A.~Martynyuk,
  ``Reconfigurable reflectarrays based on optimized spiraphase-type elements,''
  \emph{{IEEE} Trans. Antennas Propag.}, vol.~60, no.~4, pp. 1821--1830, 2012.

\bibitem{bhartia1982}
P.~Bhartia and I.~J. Bahl, ``Frequency agile microstrip antennas,''
  \emph{Microwave Journal}, pp. 67--70, Oct. 1982.

\bibitem{boccia2002_ap}
L.~Boccia, F.~Venneri, G.~Amendola, and G.~Di~Massa, ``Application of varactor
  diodes for reflectarray phase control,'' in \emph{Proc. 2002 Antennas Propag.
  Soc. Int. Symp.}, vol.~3, Jun. 2002, pp. 132--135.

\bibitem{vendik2008}
O.~Vendik and M.~Parnes, ``A phase shifter with one tunable component for a
  reflectarray antenna,'' \emph{{IEEE} Antennas Propag. Mag.}, vol.~50, no.~4,
  pp. 53--65, 2008.

\bibitem{boccia2010}
L.~Boccia, G.~Amendola, and G.~Di~Massa, ``Performance improvement for a
  varactor-loaded reflectarray element,'' \emph{{IEEE} Trans. Antennas
  Propag.}, vol.~58, no.~2, pp. 585--589, 2010.

\bibitem{legay2007}
H.~Legay, Y.~Cailloce, O.~Vendier, G.~Caille, J.~Perruisseau-Carrier, M.~Lathi,
  J.~Polizzi, U.~Oestermann, P.~Pons, and N.~Raveu, ``Satellite antennas based
  on {MEMS} tunable reflectarrays,'' in \emph{Proc. 2007 Europ. Conf. Antennas
  Propag. (EuCAP2007)}, 2007.

\bibitem{rajagopalan2008}
H.~Rajagopalan, Y.~Rahmat-Samii, and W.~Imbriale, ``{RF} {MEMS} actuated
  reconfigurable reflectarray patch-slot element,'' \emph{{IEEE} Trans.
  Antennas Propag.}, vol.~56, no.~12, pp. 3689--3699, 2008.

\bibitem{aubert2006}
H.~Aubert, N.~Raveu, E.~Perret, and H.~Legay, ``Multi-scale approach for the
  electromagnetic modelling of {MEMS}-controlled reflectarrays,'' in
  \emph{Proc. 2006 Europ. Conf. Antennas Propag. (EuCAP2006)}, 2006.

\bibitem{moessinger2006}
A.~Moessinger, R.~Marin, S.~Mueller, J.~Freese, and R.~Jakoby, ``Electronically
  reconfigurable reflectarrays with nematic liquid crystals,'' \emph{Electron.
  Lett.}, vol.~42, no.~16, pp. 899--900, 3, 2006.

\bibitem{hu2008}
W.~Hu, R.~Cahill, J.~Encinar, R.~Dickie, H.~Gamble, V.~Fusco, and N.~Grant,
  ``Design and measurement of reconfigurable millimeter wave reflectarray cells
  with nematic liquid crystal,'' \emph{{IEEE} Trans. Antennas Propag.},
  vol.~56, no.~10, pp. 3112--3117, 2008.

\bibitem{romanofsky2000}
R.~Romanofsky, J.~Bernhard, F.~Van~Keuls, F.~Miranda, G.~Washington, and
  C.~Canedy, ``K-band phased array antennas based on {Ba0.60Sr0.40TiO3}
  thin-film phase shifters,'' \emph{{IEEE} Trans. Microw. Theory Tech.},
  vol.~48, no.~12, pp. 2504--2510, 2000.

\bibitem{sazegar2009}
M.~Sazegar, A.~Giere, Y.~Zheng, H.~Maune, A.~Moessinger, and R.~Jakoby,
  ``Reconfigurable unit cell for reflectarray antenna based on
  barium-strontium-titanate thick-film ceramic,'' in \emph{2009 Europ.
  Microwave Conf. (EuMC 2009)}, 2009, pp. 598--601.

\bibitem{munk2000book}
B.~A. Munk, \emph{Frequency Selective Surfaces: Theory and Design}.\hskip 1em
  plus 0.5em minus 0.4em\relax New York: Wiley-Interscience, 2000.

\bibitem{karnati2013}
K.~Karnati, Y.~Yusuf, S.~Ebadi, and X.~Gong, ``Theoretical analysis on
  reflection properties of reflectarray unit cells using quality factors,''
  \emph{{IEEE} Trans. Antennas Propag.}, vol.~61, no.~1, pp. 201--210, 2013.

\bibitem{sievenpiper2002}
D.~Sievenpiper, J.~Schaffner, R.~Loo, G.~Tangonan, S.~Ontiveros, and R.~Harold,
  ``A tunable impedance surface performing as a reconfigurable beam steering
  reflector,'' \emph{{IEEE} Trans. Antennas Propag.}, vol.~50, no.~3, pp.
  384--390, Mar. 2002.

\bibitem{Car2012a}
E.~Carrasco, M.~Barba, and J.~Encinar, ``X-band reflectarray antenna with
  switching-beam using pin diodes and gathered elements,'' \emph{{IEEE} Trans.
  Antennas Propag.}, vol.~60, no.~12, pp. 5700--5708, 2012.

\bibitem{Deb2010e}
T.~Debogovic, J.~Perruisseau-Carrier, and J.~Bartolic, ``Partially reflective
  surface antenna with dynamic beamwidth control,'' \emph{{IEEE} Antennas
  Wireless Propag. Lett.}, vol.~9, pp. 1157--1160, 2010.

\bibitem{Bay2012a}
O.~Bayraktar, O.~Civi, and T.~Akin, ``Beam switching reflectarray
  monolithically integrated with rf mems switches,'' \emph{{IEEE} Trans.
  Antennas Propag.}, vol.~60, no.~2, pp. 854--862, 2012.

\bibitem{Car2012g}
E.~Carrasco, M.~Barba, B.~Reig, C.~Dieppedale, and J.~Encinar,
  ``Characterization of a reflectarray gathered element with electronic control
  using ohmic {RF} {MEMS} and patches aperture-coupled to a delay line,''
  \emph{{IEEE} Trans. Antennas Propag.}, vol.~60, no.~9, pp. 4190--4201, 2012.

\bibitem{Phe1977a}
H.~R. Phelan, ``Spiraphase reflectarray for multitarget radar,''
  \emph{Microwave Journal}, vol.~20, p.~67, Jul. 1977.

\bibitem{Guc2012a}
C.~Guclu, J.~Perruisseau-Carrier, and O.~Civi, ``Proof of concept of a
  dual-band circularly-polarized {RF} {MEMS} beam-switching reflectarray,''
  \emph{{IEEE} Trans. Antennas Propag.}, vol.~60, no.~11, pp. 5451--5455, 2012.

\bibitem{Phi2008a}
R.~Phillion and M.~Okoniewski, ``Improving the phase resolution of a
  micromotor-actuated phased reflectarray,'' in \emph{2008 Microsystems and
  Nanoelectronics Research Conference (MNRC 2008)}, 2008, pp. 169--172.

\bibitem{Per2010a}
J.~Perruisseau-Carrier, ``Versatile reconfiguration of radiation patterns,
  frequency and polarization: A discussion on the potential of controllable
  reflectarrays for software-defined and cognitive radio systems,'' in
  \emph{2010 IEEE International Microwave Workshop Series on RF Front-ends for
  Software Defined and Cognitive Radio Solutions (IMWS),}, 2010.

\bibitem{Per2009a}
J.~Perruisseau-Carrier and P.~Pardo, ``Unit cells for dual-polarized and
  polarization-flexible reflectarrays with scanning capabilities,'' in
  \emph{Proc. 2009 Europ. Conf. Antennas Propag. (EuCAP2009)}, 2009, pp.
  1218--1221.

\bibitem{Per2010b}
J.~Perruisseau-Carrier, ``Dual-polarized and polarization-flexible reflective
  cells with dynamic phase control,'' \emph{{IEEE} Trans. Antennas Propag.},
  vol.~58, no.~5, pp. 1494--1502, 2010.

\bibitem{Car2010a}
E.~Carrasco, J.~Encinar, M.~Barba, R.~Vincenti, and R.~Sorrentino,
  ``Dual-polarization reflectarray elements for ku-band tx/rx portable terminal
  antenna,'' in \emph{Proc. 2010 Europ. Conf. Antennas Propag. (EuCAP2010)},
  2010.

\bibitem{Car2013c}
E.~Carrasco, B.~Mariano, J.~A. Encinar, and J.~Perruisseau-Carrier, ``Two-bit
  reflectarray elements with phase and polarization reconfiguration
  (accepted),'' in \emph{Proc. 2013 IEEE Int. Symp. Antennas Propag.}, Orlando,
  Florida, 2013.

\bibitem{Men2013a}
S.~Mener, R.~Gillard, R.~Sauleau, C.~Cheymol, and P.~Potier, ``Design and
  characterization of a {CPSS}-based unit-cell for circularly polarized
  reflectarray applications,'' \emph{{IEEE} Trans. Antennas Propag.}, vol.~61,
  no.~4, pp. 2313--2318, 2013.

\bibitem{Rod2013a}
D.~Rodrigo, L.~Jofre, and J.~Perruisseau-Carrier, ``A frequency reconfigurable
  cell for beam-scanning reflectarrays (accepted),'' in \emph{Proc. 2013 IEEE
  Int. Symp. Antennas Propag.}, Orlando, Florida, 2013.

\bibitem{navarro1996book}
J.~A. Navarro and K.~Chang, \emph{Integrated Active Antennas and Spatial Power
  Combining}.\hskip 1em plus 0.5em minus 0.4em\relax John Wiley \& Sons, 1996.

\bibitem{kishor2012}
K.~Kishor and S.~Hum, ``An amplifying reconfigurable reflectarray antenna,''
  \emph{{IEEE} Trans. Antennas Propag.}, vol.~60, no.~1, pp. 197--205, 2012.

\bibitem{bialkowski2002}
M.~E. Bialkowski, A.~W. Robinson, and H.~J. Song, ``Design, development, and
  testing of {X}-band amplifying reflectarrays,'' \emph{{IEEE} Trans. Antennas
  Propag.}, vol.~50, no.~8, pp. 1065--1076, Aug. 2002.

\bibitem{clark2003}
R.~W. Clark, G.~H. Huff, and J.~T. Bernhard, ``An integrated active microstrip
  reflectarray element with an internal amplifier,'' \emph{{IEEE} Trans.
  Antennas Propag.}, vol.~51, no.~5, pp. 993--999, May 2003.

\bibitem{cabria2009}
{Lorena Cabria, Jos\'{e} \'{A}ngel Garc\'{i}a, Julio Guti\'{e}rrez-R\'{i}os,
  Antonio Taz\'{o}n, and Juan Vassal'lo}, ``Active reflectors possible
  solutions based on reflectarrays and fresnel reflectors,'' \emph{Int. Jour.
  Antennas Propag.}, vol. 2009, Article ID 653952, 13 pages, 2009.

\bibitem{ruggerini2010}
G.~Ruggerini, G.~Toso, and P.~Angeletti, ``A discrete aperiodic active lens for
  multibeam satellite applications,'' in \emph{2010 IEEE International
  Symposium on Phased Array Systems and Technology (ARRAY)}, 2010, pp.
  524--528.

\bibitem{al-joumayly2010}
M.~A. Al-Joumayly and N.~Behdad, ``A generalized method for synthesizing
  low-profile, band-pass frequency selective surfaces with non-resonant
  constituting elements,'' \emph{{IEEE} Trans. Antennas Propag.}, vol.~58,
  no.~12, pp. 4033 --4041, 2010.

\bibitem{lau2011}
J.~Lau and S.~Hum, ``Analysis and characterization of a multipole
  reconfigurable transmitarray element,'' \emph{{IEEE} Trans. Antennas
  Propag.}, vol.~59, no.~1, pp. 70--79, 2011.

\bibitem{torre2008}
P.~de~la Torre and M.~Sierra-Castaner, ``Electronically reconfigurable patches
  for transmit-array structures at 12 {GHz},'' in \emph{Proc. 2008 IEEE
  Antennas Propag. Soc. Int. Symp.}, 2008.

\bibitem{clemente2012}
A.~Clemente, L.~Dussopt, R.~Sauleau, P.~Potier, and P.~Pouliguen, ``1-bit
  reconfigurable unit cell based on {PIN} diodes for transmit-array
  applications in -band,'' \emph{{IEEE} Trans. Antennas Propag.}, vol.~60,
  no.~5, pp. 2260--2269, 2012.

\bibitem{iturri-hinojosa2010}
A.~Iturri-Hinojosa, J.~Martinez-Lopez, and A.~Martynyuk, ``Analysis and design
  of {E}-plane scanning grid arrays,'' \emph{{IEEE} Trans. Antennas Propag.},
  vol.~58, no.~7, pp. 2266 --2274, Jul. 2010.

\bibitem{boccia2011}
L.~Boccia, I.~Russo, G.~Amendola, and G.~Di~Massa, ``Preliminary results on
  tunable frequency selective surface for beam steering transmit-array
  applications,'' in \emph{Proc. 2011 Europ. Conf. Antennas Propag.
  (EuCAP2011)}, 2011, pp. 1002--1005.

\bibitem{lau2012b}
J.~Y. Lau and S.~V. Hum, ``A wideband reconfigurable transmitarray element,''
  \emph{{IEEE} Trans. Antennas Propag.}, vol.~60, no.~3, 2012.

\bibitem{jiang2012}
T.~Jiang, Z.~Wang, D.~Li, J.~Pan, B.~Zhang, J.~Huangfu, Y.~Salamin, C.~Li, and
  L.~Ran, ``Low-{DC} voltage-controlled steering-antenna radome utilizing
  tunable active metamaterial,'' \emph{{IEEE} Trans. Microw. Theory Tech.},
  vol.~60, no.~1, pp. 170--178, 2012.

\bibitem{torre2010}
P.~Padilla, A.~Mu{\~n}oz-Acevedo, M.~Sierra-Castaner, and M.~Sierra-P{\'e}rez,
  ``Electronically reconfigurable transmitarray at {Ku} band for microwave
  applications,'' \emph{{IEEE} Trans. Antennas Propag.}, vol.~58, no.~8, pp.
  2571--2579, 2010.

\bibitem{lau2012a}
J.~Lau and S.~Hum, ``A wideband reconfigurable transmitarray element,''
  \emph{{IEEE} Trans. Antennas Propag.}, vol.~60, no.~3, pp. 1303--1311, 2012.

\bibitem{cheng2009}
C.-C. Cheng, B.~Lakshminarayanan, and A.~Abbaspour-Tamijani, ``A programmable
  lens-array antenna with monolithically integrated {MEMS} switches,''
  \emph{{IEEE} Trans. Microw. Theory Tech.}, vol.~57, no.~8, pp. 1874--1884,
  Aug. 2009.

\bibitem{okane2009}
S.~O'Kane and V.~Fusco, ``Circularly polarized curl antenna lens with manual
  tilt properties,'' \emph{{IEEE} Trans. Antennas Propag.}, vol.~57, no.~12,
  pp. 3984--3987, 2009.

\bibitem{euler2010}
M.~Euler and V.~Fusco, ``Frequency selective surface using nested split ring
  slot elements as a lens with mechanically reconfigurable beam steering
  capability,'' \emph{{IEEE} Trans. Antennas Propag.}, vol.~58, no.~10, pp.
  3417--3421, 2010.

\bibitem{kaouach2011}
H.~Kaouach, L.~Dussopt, J.~Lant{\'e}ri, T.~Koleck, and R.~Sauleau, ``Wideband
  low-loss linear and circular polarization transmit-arrays in v-band,''
  \emph{{IEEE} Trans. Antennas Propag.}, vol.~59, no.~7, pp. 2513--2523, 2011.

\bibitem{liu2010}
C.~Liu and S.~Hum, ``An electronically tunable single-layer reflectarray
  antenna element with improved bandwidth,'' \emph{{IEEE} Antennas Wireless
  Propag. Lett.}, vol.~9, pp. 1241--1244, 2010.

\bibitem{perez-palomino2012}
G.~Perez-Palomino, J.~Encinar, M.~Barba, and E.~Carrasco, ``Design and
  evaluation of multi-resonant unit cells based on liquid crystals for
  reconfigurable reflectarrays,'' \emph{IEE Proc. Microwaves Antennas Propag.},
  vol.~6, no.~3, pp. 348--354, 2012.

\bibitem{carrasco2007}
E.~Carrasco, M.~Barba, and J.~Encinar, ``Reflectarray element based on
  aperture-coupled patches with slots and lines of variable length,''
  \emph{{IEEE} Trans. Antennas Propag.}, vol.~55, no.~3, pp. 820--825, 2007.

\bibitem{carrasco2011}
------, ``Design and validation of gathered elements for steerable-beam
  reflectarrays based on patches aperture-coupled to delay lines,''
  \emph{{IEEE} Trans. Antennas Propag.}, vol.~59, no.~5, pp. 1756--1761, 2011.

\bibitem{kundtz2011}
N.~Kundtz, D.~Smith, and J.~Pendry, ``Electromagnetic design with
  transformation optics,'' \emph{Proc. {IEEE}}, vol.~99, no.~10, pp.
  1622--1633, 2011.

\bibitem{tang2010}
W.~Tang, C.~Argyropoulos, E.~Kallos, W.~Song, and Y.~Hao, ``Discrete coordinate
  transformation for designing all-dielectric flat antennas,'' \emph{{IEEE}
  Trans. Antennas Propag.}, vol.~58, no.~12, pp. 3795--3804, 2010.

\bibitem{yang2011}
R.~Yang, W.~Tang, and Y.~Hao, ``A broadband zone plate lens from transformation
  optics,'' \emph{Opt. Express}, vol.~19, no.~13, pp. 12\,348--12\,355, Jun
  2011.

\bibitem{wu2012}
Q.~Wu, J.~Turpin, and D.~Werner, ``Quasi-conformal transformation
  electromagnetics enabled flat collimating lenses,'' in \emph{Proc. 2012 IEEE
  Antennas Propag. Soc. Int. Symp.}, 2012.

\bibitem{yang2011awpl}
R.~Yang, W.~Tang, and Y.~Hao, ``Wideband beam-steerable flat reflectors via
  transformation optics,'' \emph{{IEEE} Antennas Wireless Propag. Lett.},
  vol.~10, pp. 1290--1294, 2011.

\bibitem{liang2013oe}
L.~Liang and S.~V. Hum, ``Wide-angle scannable reflector design using conformal
  transformation optics,'' \emph{Opt. Express}, vol.~21, no.~2, pp. 2133--2146,
  Jan. 2013.

\bibitem{sorrentino2009}
R.~Sorrentino, R.~Gatti, and L.~Marcaccioli, ``Recent advances on millimetre
  wave reconfigurable reflectarrays,'' in \emph{Proc. 2009 Europ. Conf.
  Antennas Propag. (EuCAP2009)}, 2009, pp. 2527--2531.

\bibitem{rajagopalan2012}
H.~Rajagopalan, S.~Xu, and Y.~Rahmat-Samii, ``Reflector antenna distortion
  compensation using sub-reflectarrays: Simulations and experimental
  demonstration [{AMTA} corner],'' \emph{{IEEE} Antennas Propag. Mag.},
  vol.~54, no.~3, pp. 235--246, 2012.

\bibitem{tienda2011}
C.~Tienda, M.~Arrebola, J.~Encinar, and G.~Toso, ``Analysis of a dual-reflect
  array antenna,'' \emph{IEE Proc. Microwaves Antennas Propag.}, vol.~5,
  no.~13, pp. 1636--1645, 2011.

\bibitem{hu2009}
W.~Hu, M.~Arrebola, R.~Cahill, J.~Encinar, V.~Fusco, H.~Gamble, Y.~Alvarez, and
  F.~Las-Heras, ``94 {GHz} dual-reflector antenna with reflectarray
  subreflector,'' \emph{{IEEE} Trans. Antennas Propag.}, vol.~57, no.~10, pp.
  3043--3050, 2009.

\bibitem{farmahini-farahani2013}
M.~Farmahini-Farahani and H.~Mosallaei, ``Birefringent reflectarray metasurface
  for beam engineering in infrared,'' \emph{Opt. Lett.}, vol.~38, no.~4, pp.
  462--464, Feb 2013.

\bibitem{Niu2013a}
T.~Niu, W.~Withayachumnankul, B.~S.-Y. Ung, H.~Menekse, M.~Bhaskaran,
  S.~Sriram, and C.~Fumeaux, ``Experimental demonstration of reflectarray
  antennas at terahertz frequencies,'' \emph{Opt. Express}, vol.~21, no.~3, pp.
  2875--2889, Feb 2013.

\bibitem{Xia2012a}
X.~Chen, L.~Huang, H.~M\"{u}hlenbernd, G.~Li, B.~Bai, Q.~Tan, G.~Jin, C.-W.
  Qiu, S.~Zhang, and T.~Zentgraf, ``Dual-polarity plasmonic metalens for
  visible light,'' \emph{Nat. Commun.}, vol.~3, p. 1198, 11 2012.

\bibitem{Yu2011a}
N.~Yu, N.~Yu, P.~Genevet, M.~A. Kats, and F.~Aieta, ``Light propagation with
  phase discontinuities: Generalized laws of reflection and refraction,''
  \emph{Science}, vol. 334, no. 6054, pp. 333--337, 10 2011.

\bibitem{Lon2013a}
L.~Zou, W.~Withayachumnankul, C.~M. Shah, A.~Mitchell, M.~Bhaskaran, S.~Sriram,
  and C.~Fumeaux, ``Dielectric resonator nanoantennas at visible frequencies,''
  \emph{Opt. Express}, vol.~21, no.~1, pp. 1344--1352, Jan. 2013.

\bibitem{Tam2012a}
A.~Tamminen, J.~Ala-Laurinaho, D.~Gomes-Martins, J.~H{\"a}kli, P.~Koivisto,
  M.~K{\"a}rkk{\"a}inen, S.~M{\"a}kel{\"a}, P.~Pursula, P.~Rantakari,
  M.~Sipil{\"a}, J.~S{\"a}ily, R.~Tuovinen, M.~Varonen, K.~A.~I. Halonen,
  A.~Luukanen, and A.~V. R{\"a}is{\"a}nen, ``Reflectarray for {120-GHz} beam
  steering application: design, simulations, and measurements,'' in \emph{Proc.
  SPIE 8362, Passive and Active Millimeter-Wave Imaging XV}, May 2012, pp.
  836\,205--836\,205.

\end{thebibliography}

% Generated by IEEEtran.bst, version: 1.13 (2008/09/30)

\end{document}